# The Influence of Temperature Anisotropies in Controlling the Development of Magnetospheric Substorms.

R. M. Winglee and E. M. Harnett
Department of Earth and Space Sciences
University of Washington
Seattle, WA 98195-351310

## Abstract

Ion anisotropies can affect a host of processes within the magnetosphere, from modifying the growth rate of various instabilities to the energization and mass transport within the magnetosphere. Global multi-fluid simulations using a full treatment of the pressure tensor are used to evaluate the influence of temperature anisotropies on the magnetospheric dynamics for an idealized substorm. The simulations are able to resolve the development of conics over the polar cap which eventually turn into beams in the lobes. Entry of this plasma, particularly heavy ions with high $T_\parallel$, leads to a faster reconnection rate, additional turbulence within the reconnection region, and the substorm onset time occurs approximately 5 min earlier relative to isotropic simulations. The anisotropic treatment yields much more intense auroral currents on the nightside associated with onset and a faster expansion phase of the substorm. The plasma entering into the current sheet experiences stronger heating in the anisotropic simulations, which increases plasma transport into the equatorial dayside. Hence, the convection of the plasmasphere towards the dayside magnetosphere during the growth phase is accompanied by a relative increase in $T_\parallel$. This leads to a very distinct temperature profile on the dayside. The injection of the energetic particles from the substorm leads to an overall increase $T_\perp$ on the dayside, but only $O^+$ reaches sufficient energies for the bulk of these ions to be able to cross noon local time during the expansion phase to participate in the formation of symmetric ring current. The net effect from these processes is to decrease the cross-polar cap potential, modify the timing of the development of reconnection and the development of the auroral currents and enhance the injection of energetic particles into the inner magnetosphere.

## 1. Introduction.

Temperature anisotropies are present throughout much of the terrestrial magnetosphere, and these anisotropies can affect the heating and transport of the plasma. Important examples include the magnetosheath where the perpendicular temperature ($T_\perp$) is greater than the parallel temperature $T_\parallel$ [e.g., *Tsurutani et al.,* 1982; *Sckopke et al.* 1990]. Under these conditions, the ion cyclotron and mirror mode instabilities can develop and reduce the overall temperature anisotropy [*Gary,* 1992]. The development of these instabilities is dependent on the plasma beta [*Gary,* 1992] as well as on the plasma composition with the presence of helium ions possibly suppressing the ion cyclotron instability [*Price et al.,* 1986].

Temperature anisotropies persist within the plasma mantle with $T_\parallel/T_\perp \sim 0.5$ [*Pilipp and Morfil,* 1976]. Lobe reconnection can produce increases in this anisotropy [*Bavaasano Cattaneo et al.,* 2006]. In the lobe, highly field-aligned beams are seen [*Sharp et al.,* 1981], which can enter into the central plasma sheet [*Frank et al.,* 1996]. Using AMPTE/IRM measurements, *Baumjohann et al.* [1988] quantified the variations in the ion temperature and the level of anisotropy across the plasma sheet boundary layer (PSBL) and in the central plasma sheet (CPS). In this study, the CPS was identified by its high density and temperature while PSBL was identified by the presence of counter streaming ions with keV ions present, which distinguished it from the lobe. *Baumjohann et al.* [1988] showed that at low bulk speeds the PSBL was on average isotropic, though there could be significant swing with either $T_\parallel$ or $T_\perp$ dominating. As the bulk speed increases, excess $T_\parallel$ tended to dominate. The CPS has less variation but the same trends occurring, with $T_\parallel$ dominating as the bulk speed increases.

These anisotropies can modify the growth rate of important instabilities within the magnetosphere that control energy and mass transport within the magnetosphere. For example, temperature anisotropies are known to modify reconnection rates. In the double adiabatic model of *Hesse and Birn* [1992] anisotropies in the plasma sheet boundary layer could significantly modify the reconnection rate with the presence of $T_\perp/T_\parallel > 1$ suppressing reconnection [*Birn et al.,* 1995; *Birn and Hesse,* 2001]. On the other hand the growth rate of the ballooning mode instability is enhanced in the presence of anisotropies, particularly if $T_\perp/T_\parallel > 2$ [*Chan et al.,* 1994; *Cheng and Qian,* 1994]. The presence of temperature anisotropies can enhance the growth rate of the ion cyclotron and mirror mode instabilities, which can be important in the dynamics of the magnetosheath as well as in the inner magnetosphere, where such waves can affect the development of the ring current [*Jordanova et al.,* 2001; *Mouikis et al.,* 2002].

The presence of temperature anisotropies can also be used as a diagnostic for the sources of various particle populations and acceleration/transport processes within the magnetosphere. For example in investigations of the formation of the cold dense plasma sheet during northward interplanetary magnetic field (IMF), *Nishino et al.* [2007] using Geotail data found that on the dusk side, the plasma consisted of two populations of protons - cold solar wind ions and hot magnetospheric plasma. The cold component has its parallel temperature enhanced on the tail flank and the perpendicular component enhanced on the dayside. The hot component appeared to be nearly isotropic in the tail with an enhanced perpendicular temperature on the dayside.

More recent observations from Cluster by *Cai et al.* [2008] have been able to separate out some of the time and spatial variations of the temperature anisotropy during the formation of a thin current sheet. They note that enhancements in the proton density can occur in association with $T_\perp/T_\parallel <1$ while oxygen can enter the current sheet with $T_\perp/T_\parallel >1$, indicating that there are mass dependencies to the development of the anisotropy. As thinning of the current sheet continues, regions where $T_\perp/T_\parallel <1$ and $T_\perp/T_\parallel >1$ are observed. While the overall density of $O^+$ is small during these events, their contribution to the plasma pressure can be a significant fraction of that from the $H^+$ ions.

Such anisotropies are not just seen in association with the formation of the thin current sheets but also appear enhanced during the passage of depolarization fronts as seen by THEMIS [*Runov et al.,* 2011]. In these observations the anisotropy varies in both time and space with the largest $T_\perp/T_\parallel$ seen behind the depolarization front at distances greater than 15 $R_E$ followed by periods where $T_\perp/T_\parallel < 1$. At 10 $R_E$ the perpendicular anisotropy appears to be reduced. The source for these variations is not known.

Ring current models [see the review by *Kozyra and Liemohn,* 2003] as well as the regional tail model [*Hesse and Birn,*1992, *Birn et al.,* 1995; *Birn and Hesse,* 2001] incorporate temperature anisotropies but global models for the magnetosphere have been slower to incorporate such effects. Recently, the BATS-R-US MHD code has been modified [*Meng et al.,* 2012] to include temperature anisotropies using the double adiabatic treatment similar to the regional tail model. A potential problem with the double adiabatic treatment is that the results are strongly dependent on the isotropization factor which is needed to prevent the buildup of unphysical anisotropies. This methodology, though, is able to qualitatively reproduce some of the overall anisotropies that are seen in THEMIS data.

Part of the problem is that the double adiabatic treatment neglects the off-diagonal elements in the pressure tensor and these terms are important in controlling the redistribution of thermal energy between the different directions. Such effects have been previously noted in the comparison of hybrid simulations with double adiabatic treatments of magnetic reconnection [*Yin et al.,* 2002].

This paper details effects on the global magnetosphere from incorporating temperature anisotropies within the framework of the multi-fluid equations. In section 2 we discuss the various means of incorporating temperature anisotropies within fluid simulations. The present work uses the full pressure tensor as opposed to the reduced equations of the two temperature Chew-Goldberger-Low (CGL) equations [*Chew et al.*, 1956]. The motivation for using the full pressure tensor is that the coupling between the off-diagonal elements is self-consistently incorporated and artificial isotropization times do not have to be assumed. We also include for the first time all the major ionospheric ion species, i.e. $H^+$, $He^+$ and $O^+$, so that mass dependencies of the outflow, energization and the development of temperature anisotropies could be examined in more detail.

We demonstrate in section 3 that while the gross characteristics of the magnetosphere including the cross polar cap potential and the total magnitude of the auroral field-aligned currents have a similar appearance, the presence of temperature anisotropies can produce significant differences including (1) reduction in the cross-polar cap potential by 10-20%, (2) changing substorm onset time by several minutes, and (3) increasing the intensity of the currents around the auroral oval, including differences in both the Region I and II current systems.

In Section 4 the properties of the particle acceleration in the magnetosphere are examined. It is shown that the anisotropic treatment yields anisotropies through the magnetosphere similar to the above-cited observations without the need for including an artificial isotropization factor. We demonstrate that the temperature anisotropy treatment yields greater particle energization throughout much of the magnetosphere, including the changes in the temperature anisotropy within the PSBL and CPS as a function of substorm activity. In addition, the anisotropic treatment shows that the dayside ionospheric outflows are dominated by conics (in the sense that $T_\perp/T_\parallel \gg 1$) while the nightside outflows are dominated by ions beams. The presence of these anisotropies lead to faster reconnection rates in the magnetotail and lead to the preferential injection of heavy ions into the inner magnetosphere, which aids in the formation of asymmetric ring current.

Potential signatures that can be observed by individual spacecraft are presented in Section 5. It is also shown that these anisotropies lead to higher frequency oscillations within the reconnection region and faster dipolarization in the inner magnetosphere. Results are then compared with an event where anisotropies were observed by Cluster. The simulations yield temperature anisotropies similar to the Cluster observations with distinct differences between the light and heavy ions. F. A summary of results is given in Section 6.

## 2. Incorporation of Temperature Anisotropies

Previous attempts to include temperature anisotropies into simulations have typically used the double adiabatic treatment. The double adiabatic model uses the CGL equations [*Chew et al.*, 1956] which have the form:

$$\frac{\partial P_\parallel}{\partial t} = -(\mathbf{V}\cdot\nabla)P_\parallel - P_\parallel \nabla\cdot\mathbf{V} - 2P_\parallel \hat{\mathbf{b}}\cdot(\hat{\mathbf{b}}\cdot\nabla)\mathbf{V} \qquad (1)$$

$$\frac{\partial P_\perp}{\partial t} = -(\mathbf{V}\cdot\nabla)P_\perp - 2P_\perp \nabla\cdot\mathbf{V} + P_\perp \hat{\mathbf{b}}\cdot(\hat{\mathbf{b}}\cdot\nabla)\mathbf{V} \qquad (2)$$

where $P_\perp$ and $P_\parallel$ are the perpendicular and parallel pressure components, $V$ is the bulk velocity of the plasma and $\hat{b} = \mathbf{B}/|\mathbf{B}|$. Note that in this treatment there are no off-diagonal elements to the pressure tensor. To minimize the development of extreme anisotropies, an isotropization factor is included in the above equations to produced enhanced coupling between $P_\perp$ and $P_\parallel$ [*Birn et al.,* 1995]. For a regional model this may be appropriate if there are not substantial variations in the plasma conditions. However, for a global magnetospheric model there are orders of magnitude variations in magnetic field strength, density and plasma composition so that the assumption of a single isotropization factor has the potential for producing significant errors, as well as introducing an arbitrary factor that is not linked to physical processes.

There is also a physical reason why the double adiabatic model is not appropriate for the magnetosphere. This model implicitly assumes that convective effects on the pressure are weak and that the off-diagonal elements of the pressure tensor are negligible. The reality though is that the off-diagonal elements are important in determining the flow of thermal energy between the different directions and should therefore not be neglected. One consequence is seen in the full treatment is that the two perpendicular temperatures need not be necessarily the same. Such a feature is consistent with the observations of *Cai et al.* [2008], particularly for the heavy ions during thinning of the current sheet. As shown in the following, the coupling between the different components of the pressure tensor is mass dependent and therefore the assumption of a single isotropization timescale in the double adiabatic treatments is therefore problematic.

In order to overcome these issues, we have revisited the original derivation of the temperature anisotropies from the Vlasov equation. The tensor form used here was originally derived by *Chapman and Cowling* [1952], and rederived by *Siscoe* [1983]. Specifically, the evolution of the distribution function, *f*, for all collisionless plasmas is governed by the Vlasov equation:

$$\frac{\partial f}{\partial t} + \mathbf{v} \cdot \frac{\partial f}{\partial \mathbf{x}} + \mathbf{a} \cdot \frac{\partial f}{\partial \mathbf{v}} = 0 \tag{3}$$

$$\mathbf{a} = \frac{q}{m}[\mathbf{E} + \mathbf{v} \times \mathbf{B}] \tag{4}$$

where **v** and **a** are the particle velocity and the acceleration of the particle, respectively. By definition the density *n*, and pressure tensor $P_{ij}$ are given by

$$\int f d^3 \mathbf{v} = n(x,t) \tag{5}$$

$$\int \mathbf{v} f d^3 \mathbf{v} = \mathbf{V}(x,t) n(x,t) \tag{6}$$

$$\int w_i w_j f d^3 \mathbf{v} = P_{ij} \tag{7}$$

where **v** is the velocity of each individual particle and **w**= **V**-**v** is the particle thermal velocity.

The multi-fluid approach, like hybrid models, treats the electrons as a fluid. Assuming the electrons are in drift motion (i.e. no acceleration) and that the electrons have an isotropic temperature, the electron properties are described by the modified Ohm's law and pressure equation

$$\mathbf{E} + \mathbf{V}_e \times \mathbf{B} + \frac{\nabla P_e}{e n_e} = 0 \tag{8}$$

$$\frac{\partial P_e}{\partial t} = -\gamma \nabla \cdot (P_e \mathbf{V}_e) + (\gamma - 1)\mathbf{V}_e \cdot \nabla P_e \qquad (9)$$

where γ is the polytropic index. The magnetic field is given by

$$\frac{\partial \mathbf{B}}{\partial t} + \nabla \times \mathbf{E} = 0 \qquad (10)$$

Assuming quasi-neutrality, one can derive the remaining equations

$$n_e = \sum_i \frac{q_i n_i}{e}, \quad \mathbf{V}_e = \sum_i \frac{q_i n_i}{e n_e}\mathbf{V}_i - \frac{\mathbf{J}}{e n_e}, \quad \mathbf{J} = \frac{1}{\mu_0}\nabla \times \mathbf{B} \qquad (11)$$

Substitution of (11) into (8) yields the standard form of the modified Ohm's law:

$$\mathbf{E} = -\sum_i \frac{q_i n_i}{e n_e}\mathbf{V}_i \times \mathbf{B} + \frac{\mathbf{J} \times \mathbf{B}}{e n_e} - \frac{1}{e n_e}\nabla P_e \qquad (12)$$

In the multi-fluid codes the bulk moments are first derived using (5) and (6) and then simplifying approximations on (7) the various pressure equations. Taking the zeroth (i.e. $\int d^3v$) and first moments (i.e., $\int d^3v v$) of (5) yields the equations for each ion species α for the mass density $\rho_\alpha$ and bulk velocity $\mathbf{V}_\alpha$

$$\frac{\partial \rho_\alpha}{\partial t} + \nabla \cdot (\rho_\alpha \mathbf{V}_\alpha) = 0 \qquad (13)$$

$$\rho_\alpha \frac{d\mathbf{V}_\alpha}{dt} = q_\alpha n_\alpha (\mathbf{E} + \mathbf{V}_\alpha \times \mathbf{B}) - \nabla \cdot \mathbf{P}_\alpha \qquad (14)$$

where $\mathbf{P}_\alpha$ is the full pressure tensor. The pressure tensor is related to the temperature tensor such that $\mathbf{P}_\alpha = n\, k\, \mathbf{P}_\alpha$ where k is Boltzmann's constant. The 2nd moment yields the pressure tensor equation given by [*Chapman and Cowling,* 1952; *Siscoe,* 1983]:

$$\frac{\partial \mathbf{P}_\alpha}{\partial t} = -\mathbf{V}_\alpha \cdot \nabla \mathbf{P}_\alpha - \mathbf{P}_\alpha \nabla \cdot \mathbf{V}_\alpha - \mathbf{P}_\alpha \cdot \nabla \mathbf{V}_\alpha - (\mathbf{P}_\alpha \cdot \nabla \mathbf{V}_\alpha)^T + \frac{e}{m_\alpha}[\mathbf{P}_\alpha \times \mathbf{B} + (\mathbf{P}_\alpha \times \mathbf{B})^T] - \nabla \cdot Q \qquad (15)$$

where $^T$ is the trace and Q is the heat flux term. For this treatment, the heat flux term is neglected, similar to the double adiabatic treatments. This is a reasonable assumption if the system is strongly controlled by convection, as in the global magnetosphere. All the other elements of the pressure tensor are retained which means that the total number of equations for each ion species consists of one density equation, three velocity equations and 6 pressure equations. This set of equations requires substantially more computer resources than the total of 5 equations for the plasma dynamics of MHD and the total of 6 equations used in the double adiabatic treatment.

The isotropic pressure equations can be derived from the above equations by assuming $P_{ij} = 0$ if i≠j and $P_{ij} = P$ for i =j, while the CGL equations can be derived by assuming

$$P_{ij} = P_\parallel \hat{b}_i \hat{b}_j + P_\perp (\delta_{ij} - \hat{b}_i \hat{b}_k) \ . \qquad (16)$$

In order to test for the effects produced by the presence of temperature anisotropies we first run the code in its isotropic form using the above averaging at each time step as well as the full anisotropic form with no averaging. Note that there are two physically distinct perpendicular directions : (i) the $V \times B$ directions and (ii) the $B \times (V \times B)$ direction. The latter is approximately the $E \times B$ direction and the temperature gains along these directions can become distinct when the scale length of a region becomes comparable to an ion gyro-radius. In order to compare the results with the double adiabatic treatments we define $P_\perp = (P_{V \times B} + P_{B \times (V \times B)})/2$ as the average of the two perpendicular pressures.

The above equations are solved using the same numerical scheme as in the multi-fluid treatments of *Winglee et al.* [2008a,b, 2011]. The equations are solved on a stacked set of 5 grid boxes. The inner most box has a grid resolution of 0.15 $R_E$ (where $R_E$ is the radius of the Earth and is equal to 6378 km) and extends from 12 $R_E$ on the day side to 30 $R_E$ in the tail, and 8 $R_E$ in each direction on the flanks. The inner boundary is set at 2.7 $R_E$. The grid spacing doubles between successive boxes with the outer most box having a grid resolution of 2.4 $R_E$ extending 50 $R_E$ on the dayside and 430 $R_E$ down tail and 120 $R_E$ on the flanks.

The density on the inner boundary is set at 150 cm$^{-3}$, which consists of 88% H$^+$, 5% He$^+$ and 7% O$^+$. This total density is lower than in previous simulations but there are more heavy ions present. The assumption of a reduced density was made since we are evaluating and isolated substorms with in a quiet time. The resistivity model at the inner boundary is the same as in previous models with the region within the inner boundary given a finite resistance equivalent to a Reynolds number of 10. At the actual inner boundary (representing the ionosphere), the Reynolds number is increased to 20 and at one grid point above it is set at 40. At all other points within the simulation grid, the resistivity is zero. These values yield an overall height integrated resistivity of about 10$^5$ Ohm m$^2$, which is of the order of the height integrated resistivity of the ionosphere at about 100 km [*Kelley*, 1989]. The corresponding resistance over the auroral zone yields approximately 0.15 mOhm. The temperature profile around the inner boundary is also similar to previously published versions of the multi-fluid model. At the equator, a high temperature of 20 eV is assumed. The plasma temperature then decreases with increasing latitude, reaching a minimum of 0.1 eV at the poles. The high temperature at the equator represents the hotter plasma trapped on closed field lines, while the low temperature over the polar cap is typical for the high latitude ionosphere.

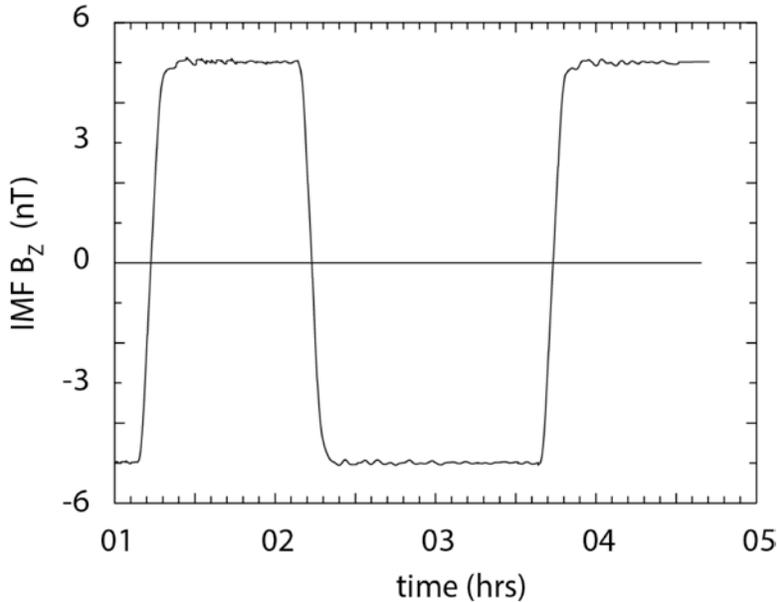

**Figure 1.** IMF conditions used to drive the simulations to test for the magnetospheric differences produced by temperature anisotropy effects.

The solar wind conditions for the simulations have a constant density of 6 cm$^{-3}$ and a speed of 450 km/s. The simulation is run initially in the isotropic mode for 2 hrs real time with zero interplanetary magnetic field (IMF) to establish an approximate equilibrium for the magnetosphere. This point in time is referred as t= 0000. The model was then run in both its isotropic and anisotropic form for another 2.5 hrs through a substorm sequence with the IMF having a value of -5 nT for 1.25 hrs and then +5 nT for another 1.25 hrs. This period allows the magnetosphere to come into equilibrium with the mass and energy transport associated with anisotropic temperature effects. At this point we then ran both types of simulations through another IMF turning as shown in Figure 1 to investigate the characteristics of the substorms as seen in the isotropic version versus the anisotropic version. The next sections detail the evolution of the substorm as seen by these two different treatments and it is shown that there are significant differences with the anisotropic treatment producing earlier tail reconnection, a more intense signature in the nightside auroral region, and stronger injection of energetic particles into the inner magnetosphere to aid in the formation of a symmetric ring current.

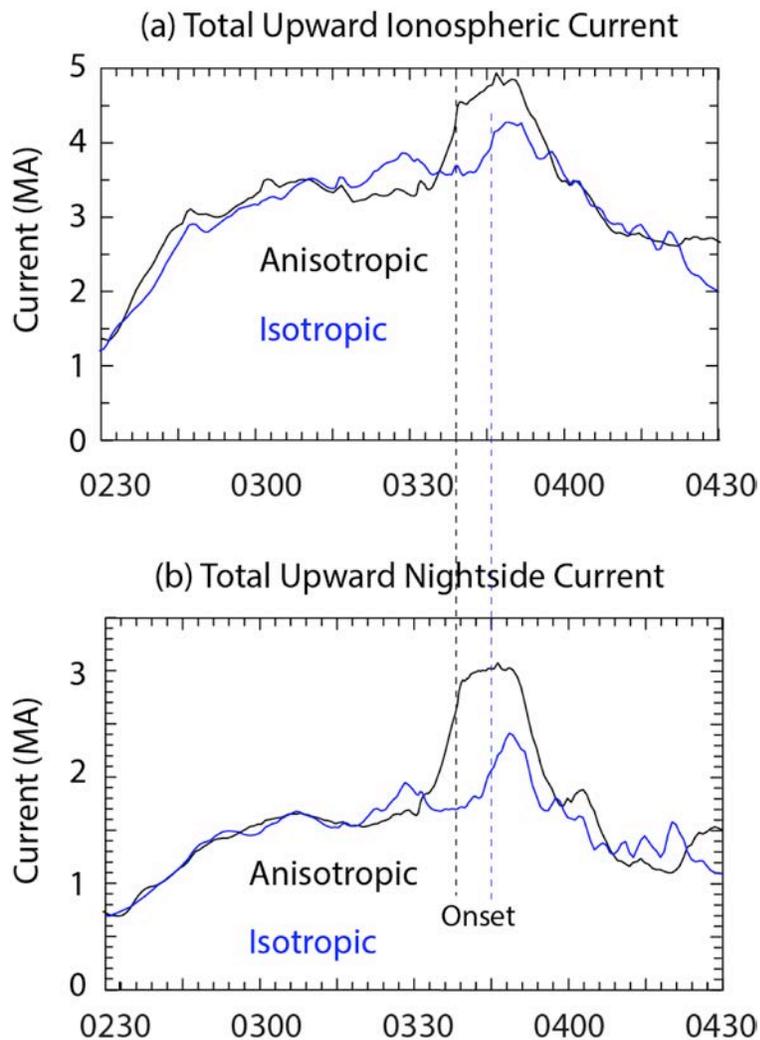

**Figure 2.** (a) Total ionospheric current into northern hemisphere and (b) contributions from the nightside region only for both the anisotropic (black) anisotropic (blue) cases. While there is an enhancement of the nightside currents associated with substorm onset for both cases, intensification is very much stronger for the anisotropic simulations.

**3. Global Response and Auroral Dynamics.**

In order to demonstrate the importance of including anisotropies into the global simulation relative to substorm processes, Figure 2 shows the evolution of (a) the total upward currents into the northern auroral region, and (b) the currents into the nightside auroral region only. It is seen that the total current for both types of simulations closely match each other through the early growth phase up to about 0330 (1 hr after the southward turning). For the anisotropic simulation, the total upward current into the ionosphere starts to rise sharply between 0330 and 0340 and this is primarily due to increases in the nightside current. These nightside current continues to rise, albeit more slowly, for another 12 mins, after which it subsides to values close to the magnitude of the total growth phase currents by about 0415. While the definition of onset can be ambiguous within the simulations we will take t = 0340 as onset time which as shown in the following corresponds to an intensification of the nightside auroral currents which subsequently expand poleward and westward in association with dipolariza-tion of the tail magnetic field (Section 5).

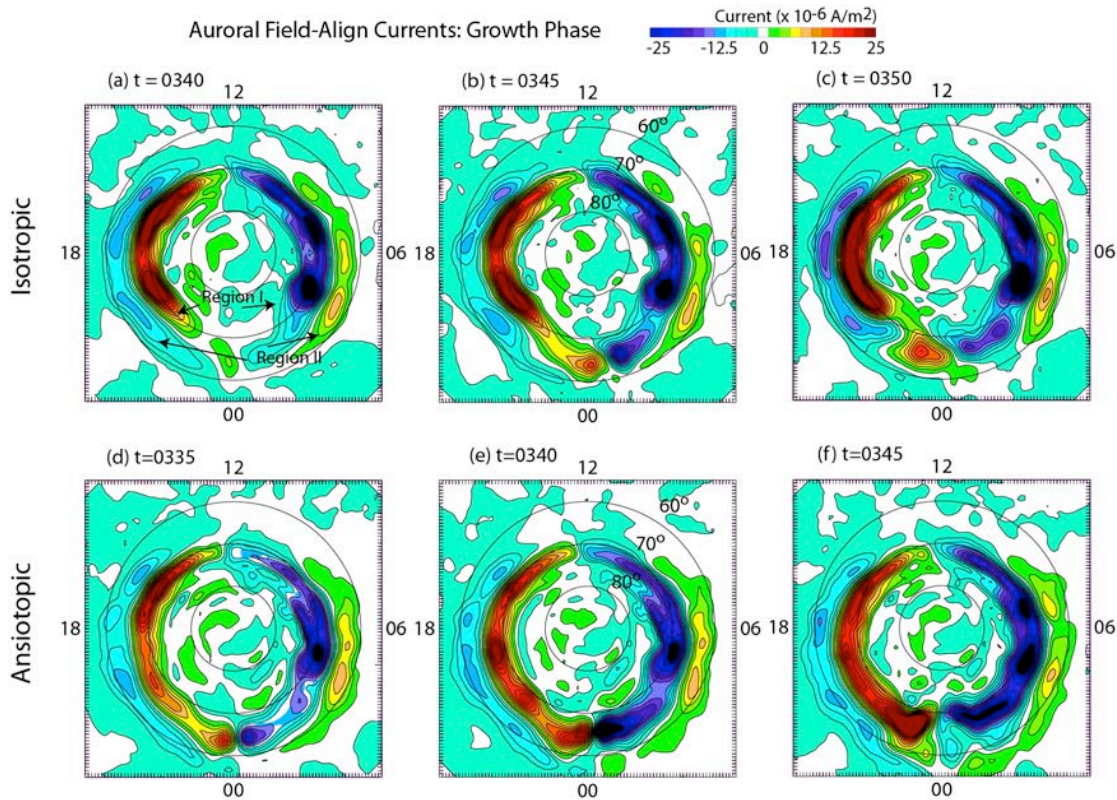

**Figure 3.** Evolution of the Region I and Region II for the northern hemisphere and for isotropic simulations (left) and anisotropic simulations (right). The panels for the isotropic simulations are 5 min behind that of the anisotropic simulations due to the delayed timing of onset between the two cases. While the overall current patterns are similar, the anisotropic simulations predict much higher currents in the nightside at onset, which is closer to the actual characteristics of auroral substorms.

For the isotropic simulation, it is seen that the enhancement of the nightside currents still occurs but it is much weaker in terms of magnitude and duration. In addition, the rapid increase in the nightside auroral currents occurs later than the anisotropic simulation. This difference in timing and magnitude as shown in the following are due to the changes in mass transport, particularly for the heavy ions, that is produced when anisotropy temperature effects are included.

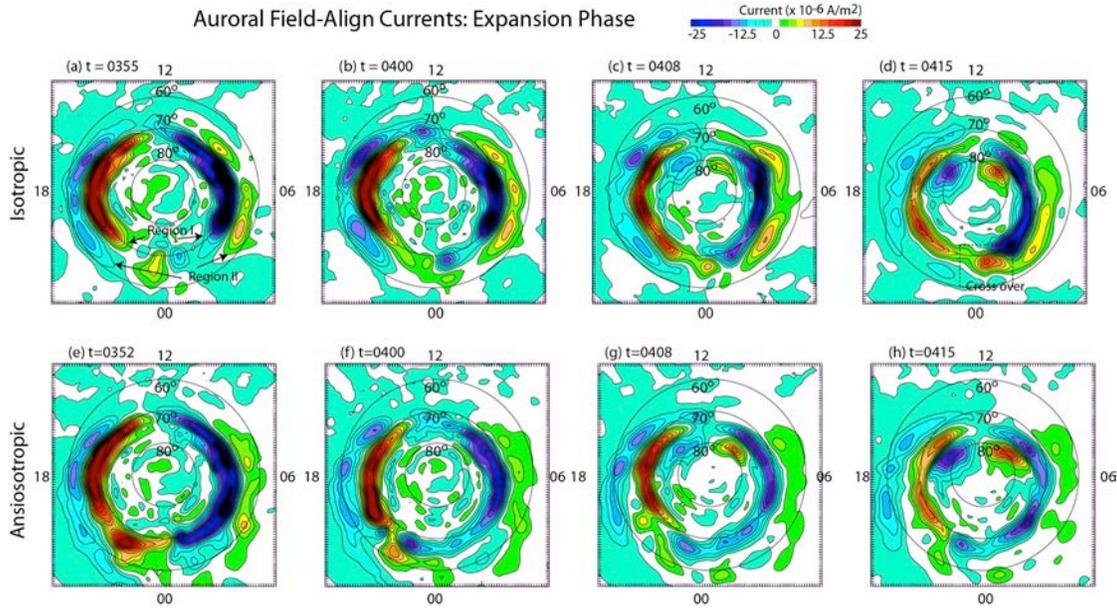

**Figure 4.** Continuation of the Figure 3. The anisotropic simulations show stronger equatorward and westward movement of nightside intensification and evolves more quickly than the isotropic simulations.

In order to more fully understand the origin of this difference, Figure 3 shows the evolution of the currents about the auroral region into the northern hemisphere. The Region I currents are on the higher latitude side and Region II currents are on the lower latitude side. . The currents are calculated five grid points (0.75 $R_E$) away from the inner boundary (so that influences of boundary effects are minimized) and then mapped to the surface of the Earth to produce the current density in absolute units in the topside ionosphere. The results for the isotropic run are shown on the left hand side and that for the anisotropy run on the right hand side. Note that the timescale has been shifted so the results are more aligned with their respective onset times.

For both simulations, onset occurs close to midnight local time at a latitude of about 65° and is associated with an enhancement of the nightside currents. For the isotropic case, the nightside enhancements remain weaker than the currents seen near the terminator. This effect reproduces the results of the earlier isotropic multi-fluid simulations of *Winglee et al.* [2004] which showed that the introduction of ion skin depth/ion cyclotron effects in the multi-fluid treatment gave higher currents in the nightside region than single fluid MHD treatments but these currents were still weaker than currents at the terminator.

For the anisotropic simulations, this is no longer the case and the nightside currents become the dominant currents near onset. These currents quickly move poleward and westward after onset, and much more rapidly than the isotropic simulations. The larger magnitude of the nightside current and the above fast movement in latitude and local time for the anisotropic simulations provide a more realistic treatment of the canonical development of substorm.

Another difference between the anisotropic and isotropic simulations is that the Region II currents on the dayside at onset are more intense for the isotropic simulations while the nightside

Region II currents are more intense for the anisotropic simulations. This difference in the currents will be demonstrated in the following section to be due to differences in the injection of energetic particles into the inner magnetosphere at onset.

Figure 4 shows the continuation of the development of the auroral currents through the expansion phase. It is seen that the nightside currents never for the isotropic simulations and reach comparable magnitudes to the peak dayside currents seen during the growth phase, though there is continued activity near midnight out to 0415. The Region II currents reach their peak value on the dayside at 0400 and there is a crossover between the dayside Region I and Region II currents at this time. A crossover of the nightside Region I and II currents occurs later at 0415. Such cross-over between these currents systems have been noted by *Iijima and Potemra* [1976].

For the continuation of the anisotropic case, the nightside currents remain strong but only for a relatively short period and are essentially gone by 0400. This difference signifies that the total energy involved is about the same between the two simulations but the anisotropic simulations lead to faster energy dissipation than the isotropic simulations.

The cross-over of the dayside Region I and II currents occurs at about the same time as the isotropic simulations at 0400. However, the cross-over of the nightside Region II currents does not occur. Instead the Region II currents for the anisotropic simulations extend to lower latitudes (<60°) indicating deeper penetration of energetic particles into the inner magnetosphere than the isotropic simulations. For a crossover of the nightside currents, the present results suggest that other processes such as a IMF $B_y$ is needed to produce this effect.

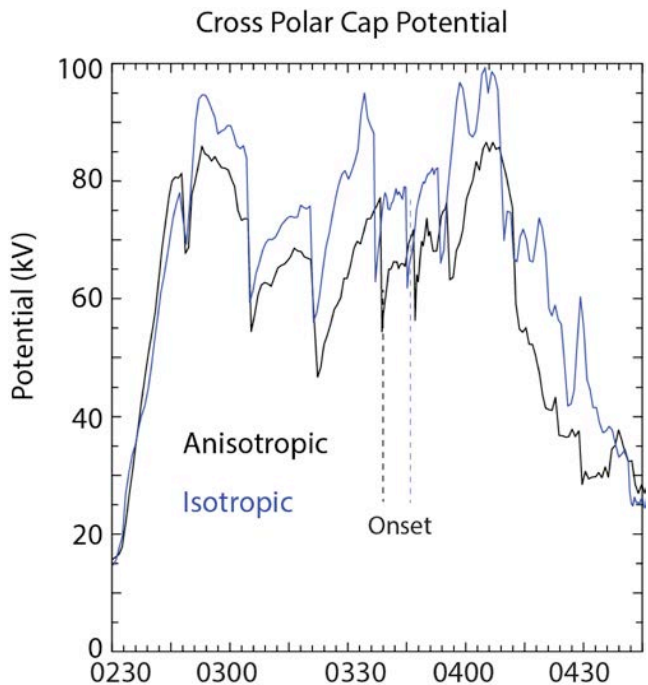

**Figure 5.** Evolution of the cross-polar cap potential for the two cases. Onset for both cases is associated with a drop in the polar cap potential but overall the polar cap potential is lower in the anisotropic case by about 10 to 20% relative to the isotropic simulations.

The evolution of the cross polar cap potential for the two cases is shown in Figure 5. It is seen that during the early growth phase the cross polar cap potential is approximately the same for the two different simulations (i.e. when the magnetosphere is directly driven by the solar wind). However, the potential for the anisotropic case has a lower saturation value by about 10% and this reduction is seen throughout the rest of the period of southward IMF. At times the potential for the anisotropic case can be as much as 20% less than the value for the isotropic simulations. The two potentials again become comparable only after the presence of a northward IMF for a few tens of minutes.

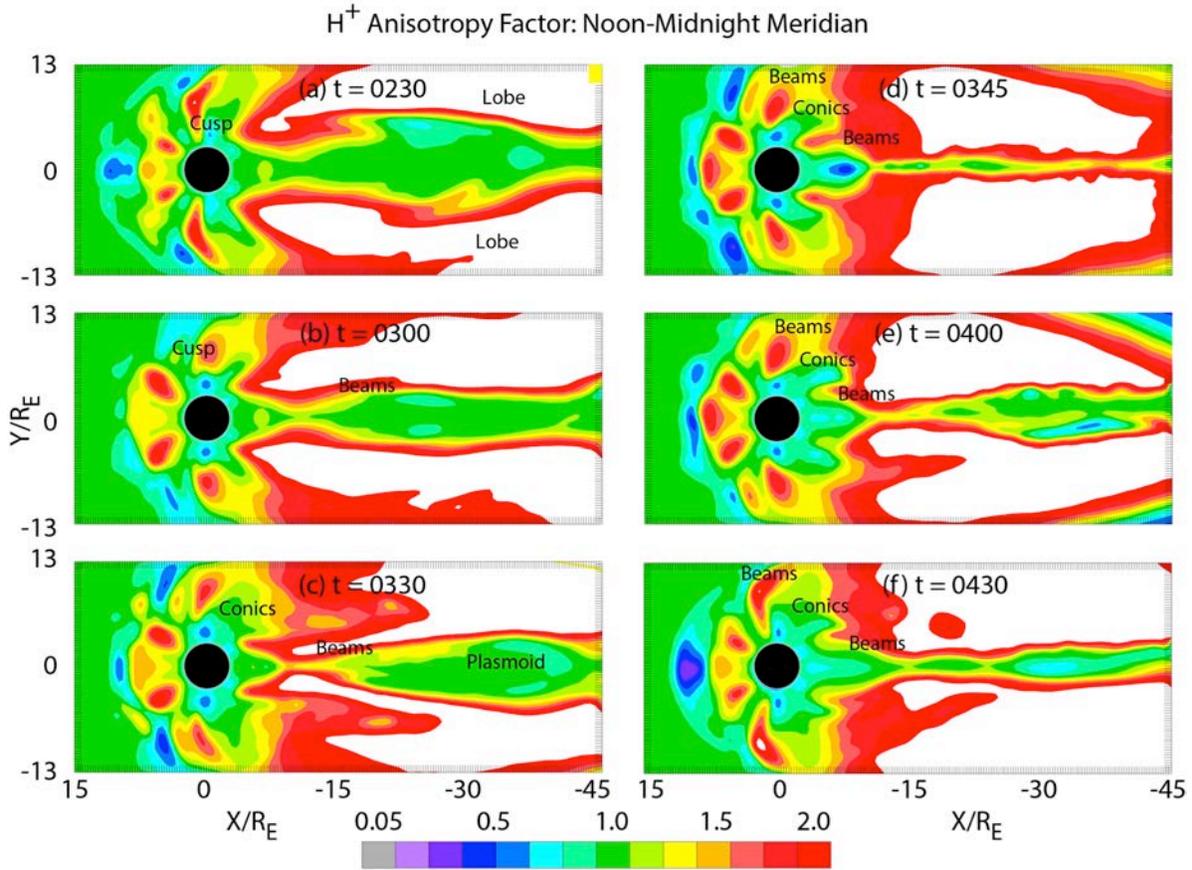

**Figure 6.** The development of the anisotropy factor $A = T_{\parallel}/T_{Av}$ for H$^+$ from the start of the growth phase into the expansion phase in the noon-midnight meridian. The lobe is seen as having a high $T_{\parallel}$ while conics are present near the cusp and over the polar cap region, with beams associated with the nightside auroral region and dayside refilling of the plasmasphere. The current sheet has a variable anisotropic factor associated with its thinning, ejection of a plasmoid and substorm activity.

As noted in *Winglee et al.*, [2002, 2004], The cross-polar cap potential tends to overestimate the observed value, though the presence of heavy ions leads to the saturation of the potential at significantly lower values. The present results indicate that a further reduction in the cross-polar cap potential can occur when temperature anisotropies are included.

**4. Development of Temperature Anisotropies and Changes to the Tail Dynamics.**

In order to show how the presence of temperature anisotropies can modify the substorm dynamics, Figure 6 shows the anisotropy factor (defined here as $A = T_{\parallel}/T_{Av}$ where $T_{Av} = (T_{\parallel} + T_{BxVxB} + T_{VxB})/3$) for H$^+$ ions in the noon-midnight meridian. The cusp is seen as a region where the anisotropy factor is less than 1 at high latitudes on the dayside as particles moving into stronger magnetic field increase their perpendicular energy at the expense of their parallel energy. Equatorward of the cusp is a region of high $T_{\parallel}$ and this region is associated with plasmaspheric refilling. Poleward of the cusp is another region of elevated $T_{\parallel}$ which is associated with mirroring magnetosheath plasma and ionospheric outflow. Over the polar cap, the ionospheric outflow has $T_\perp > T_\parallel$ which is essentially a conic (albeit within the simulations the conics are not folded over in pitch angle). It is interesting to note that the conics here are produced by convection effects and while wave-particle interactions are potentially important for conic formation they are not necessarily required. Moving out from the nightside auroral regions and into the lobes, the H$^+$ ions have $T_\perp \ll T_\parallel$, consistent with the observations of *Sharp et al.* [1981].

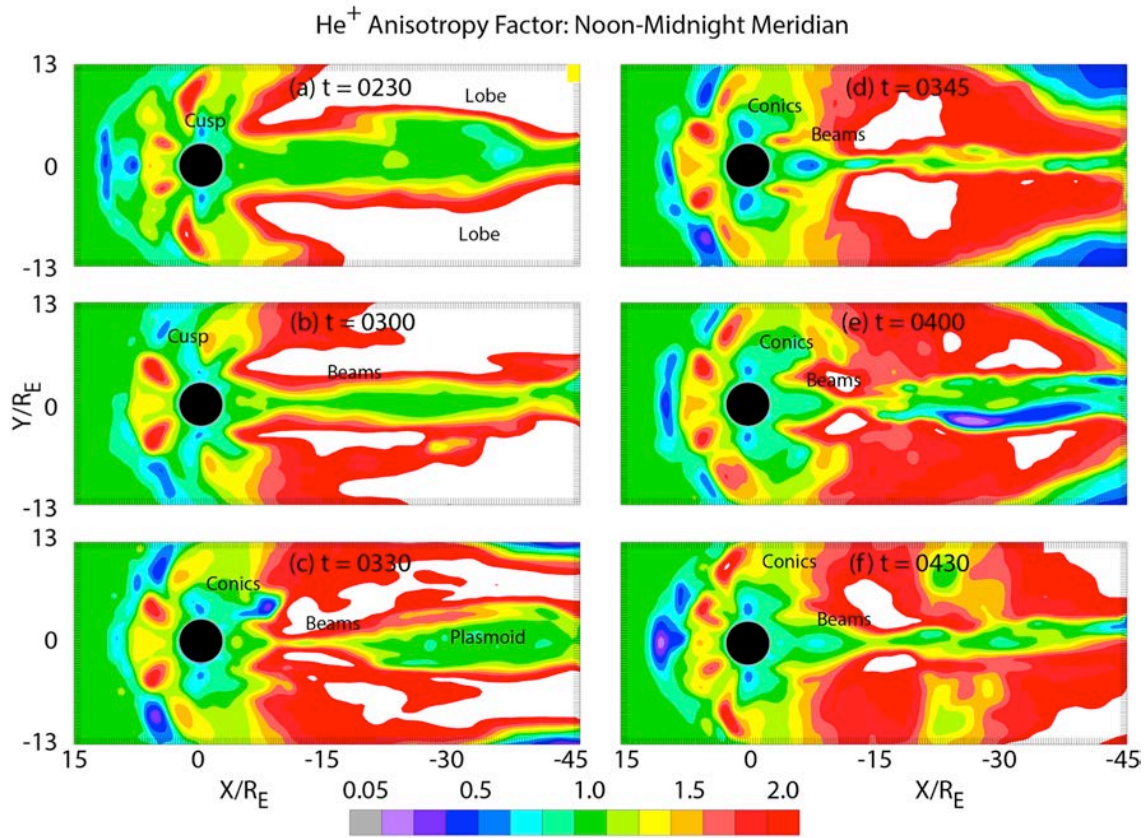

**Figure 7.** As in Figure 6 except for the He$^+$ ions. The anisotropy factor in the lobes is smaller for these intermediate ions and the conics penetrates deeper into the tail.

The plasma sheet is seen as a region of no net anisotropy ($A \approx 1$) during the early part of the growth phase. As lobe plasma is convected into the plasma sheet, regions of high $T_\parallel$ can be seen entering the central plasma sheet (e.g. at 0300 near X =- 15 R$_E$). A plasmoid is then seen to be ejected at t = 0330 which is before substorm onset (consistent with previous results from the multi-fluid simulations [*Winglee et al.*, 2011]). The plasmoids leading edge has $T_\perp > T_\parallel$ while in its wake, regions of enhanced $T_\parallel$ are seen entering the plasma sheet (t = 0345 and beyond). These variations with the evolution of the plasma sheet could explain the observed variations in $T_\perp$ and $T_\parallel$ noted by *Baumjohann et al.,* [1988].

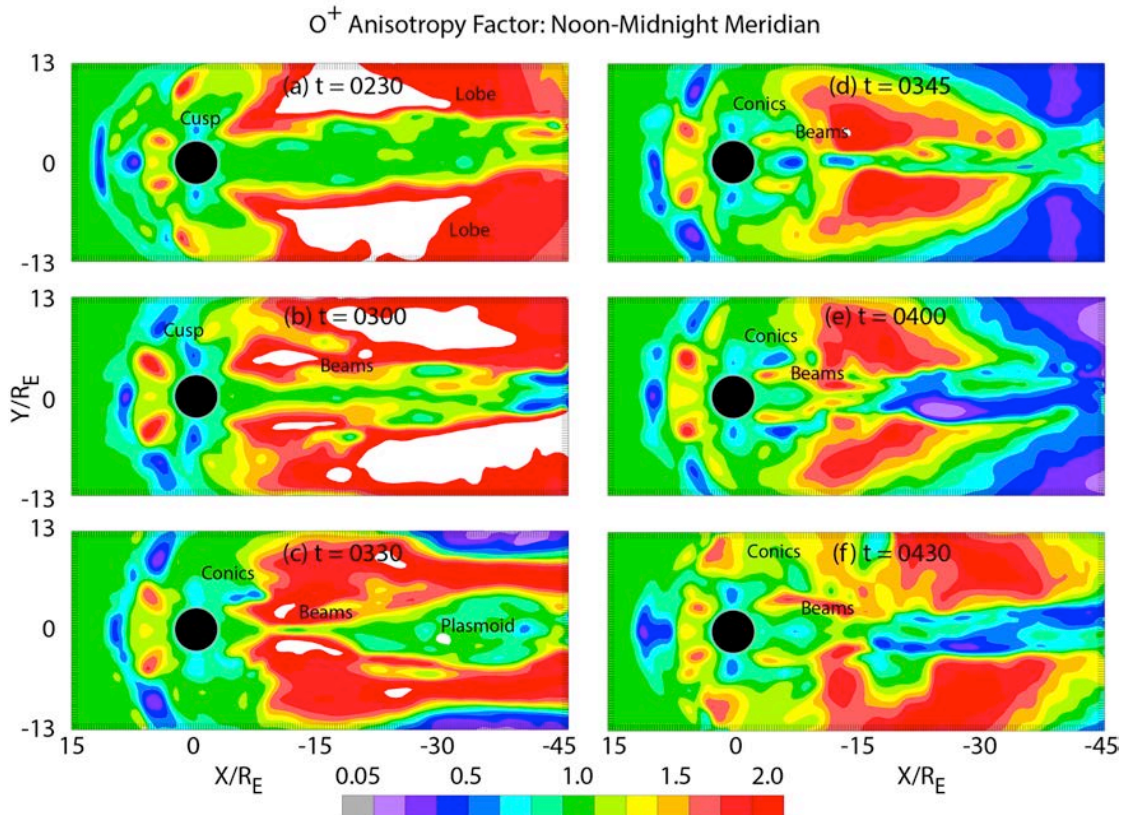

**Figure 8.** As in Figure 6 except for the O$^+$ ions. The trend of reduced $T_\parallel$ for the heavier ions continues and elevated $T_\perp$ regions are more strongly evident in the plasma sheet and inner magnetosphere.

The corresponding results for He$^+$ are shown in Figure 7. It is seen that while the overall positions for the various temperature anisotropies are the same as in Figure 6 the magnitude of the anisotropies are very different. For example, the regions of high $T_\parallel$ (A> 1) in the lobe are much smaller and do not reach as far down the tail in comparison with the H$^+$ ions. In addition, the conics (A < 1) in the polar outflows have a larger $T_\perp$ at higher altitudes and deeper into the tail than the H$^+$ ions. Another difference is that with the ejection of the plasmoid and substorm onset, populations of high $T_\parallel$ are seen to be injected into the plasma sheet.

This trend of mass dependent variations in the anisotropy factor continues for the O$^+$ ions as seen in Figure 8. The regions of high $T_\parallel$ in the lobes are the smallest for the O$^+$ ions while the polar cap conics have even access deeper penetration into the magnetotail, particularly after substorm onset. Entry of the heavy ions into the plasma sheet near onset coincides with entry injection of plasma with high $T_\parallel$ which increases the reconnection growth rate, as noted in the introduction.

The perpendicular temperature over the polar cap region is due to the influence of the convective electric field that is producing centrifugal acceleration of the ionospheric ions. In the isotropic treatment, this energization is automatically distributed in all three directions. In the anisotropic treatment there is a competition between conversion of perpendicular energy to parallel energy as the ions move into weaker magnetic field and the continual adding in of perpendicular energy through continued convection into the nightside. For the light ions, the motion down the field line is sufficiently fast that the energy conversion to parallel energy dominates. For the heavy ions, the propagation speed down the field lines is significantly slower so that the convective electric field is able to add more perpendicular temperature as the ions move into the tail.

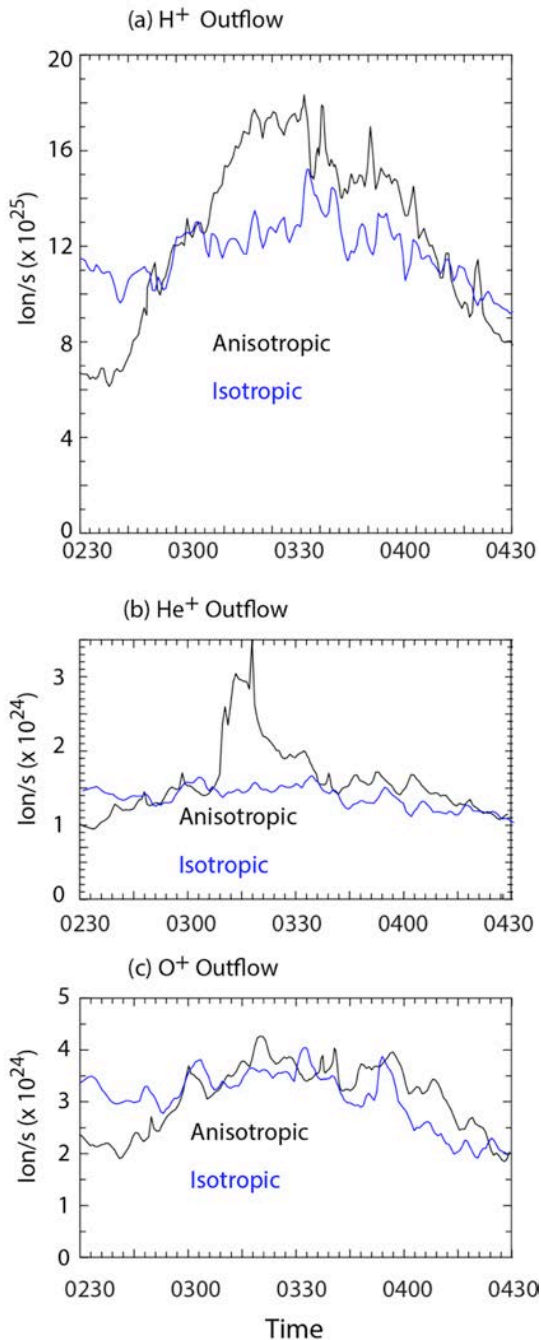

**Figure 9.** Comparison of results from the isotropic and anisotropic simulations for the total outflow rate for the different ion species. The light ions have the largest total modification to their outflow rate while the intermediate ions have the largest percent change though limited over a smaller time

The presence of these anisotropies has multiple effects on the mass and energy transport throughout the magnetosphere. One such effect is that the total outflow rate from the ionosphere is modified as shown in Figure 9. For the light ions in the anisotropic simulations, enhanced ionospheric outflows develop after about 30 min after the start of the growth phase and they continue until about 20 min after substorm onset. For the intermediate mass ions (Figure 9b), the period of enhanced ionospheric appears about 45 min after the start of the growth phase through to about substorm onset. The peak enhancement is nearly a factor of two larger than the isotropic simulations. For the heavy ions (Figure 9c) though, the outflow rates tend to be about the same during the development of the growth phase and through the expansion phase of the substorm.

Not only can the presence of temperature anisotropies modify the magnitude of the outflows, they also modify where these outflows map into the magnetotail. This effect is illustrated in Figures 10, which show the relative density of $O^+$ for the isotropic and anisotropic simulations through the event. It is seen that the outflows from the dayside map to higher latitudes in the tail while the nightside source maps to lower latitudes for the anisotropic case. This difference in mapping leads to higher heavy ion concentrations in the plasma sheet prior to substorm onset.

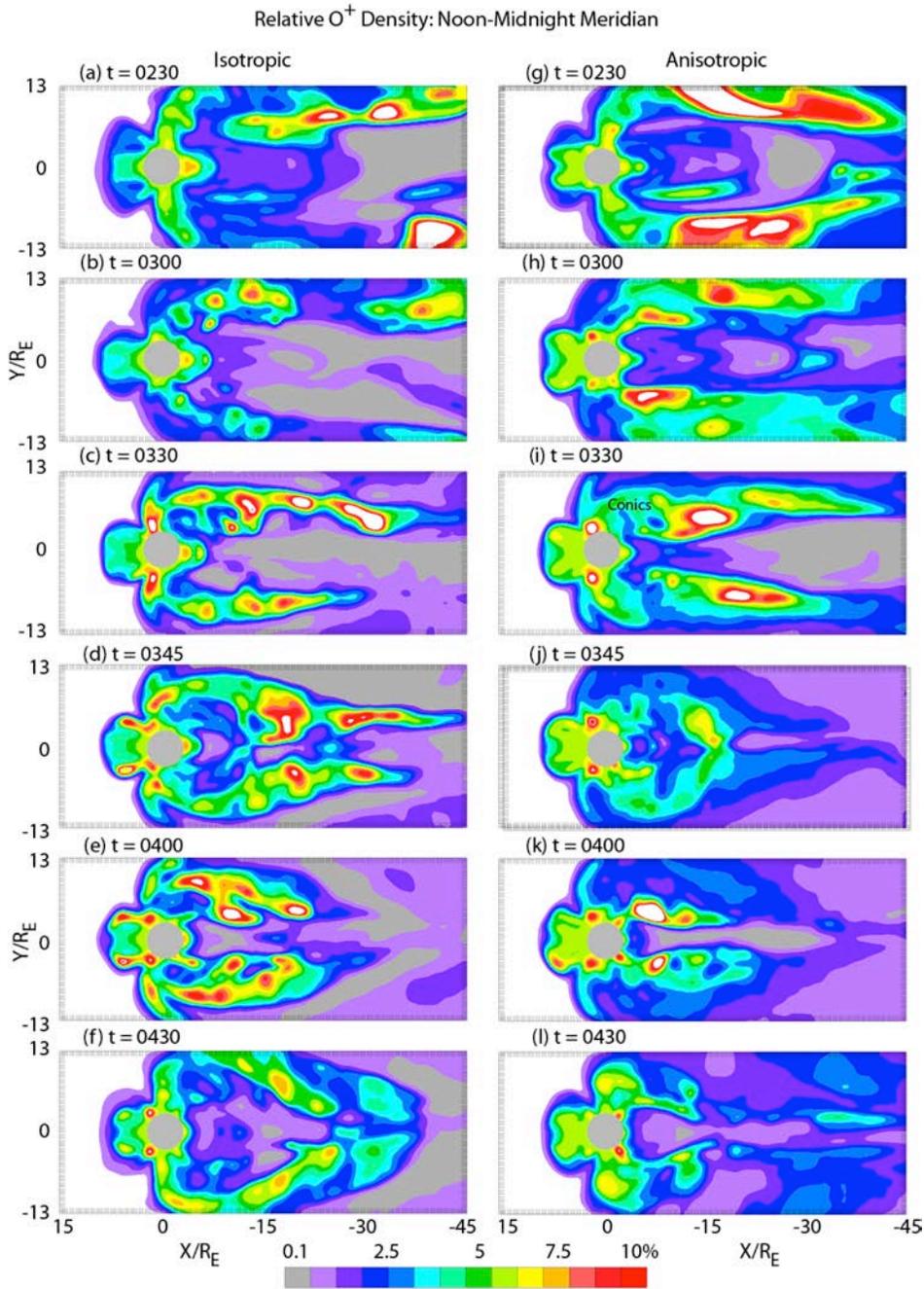

**Figure 10.** A comparison between the isotopic simulations (left hand side) and the anisotropic simulations (right hand side) of the relative density of $O^+$ in the noon-midnight meridian. The anisotropic version has a faster injection of heavy ions into the plasma sheet and subsequently lower densities in the tail during the expansion phase.

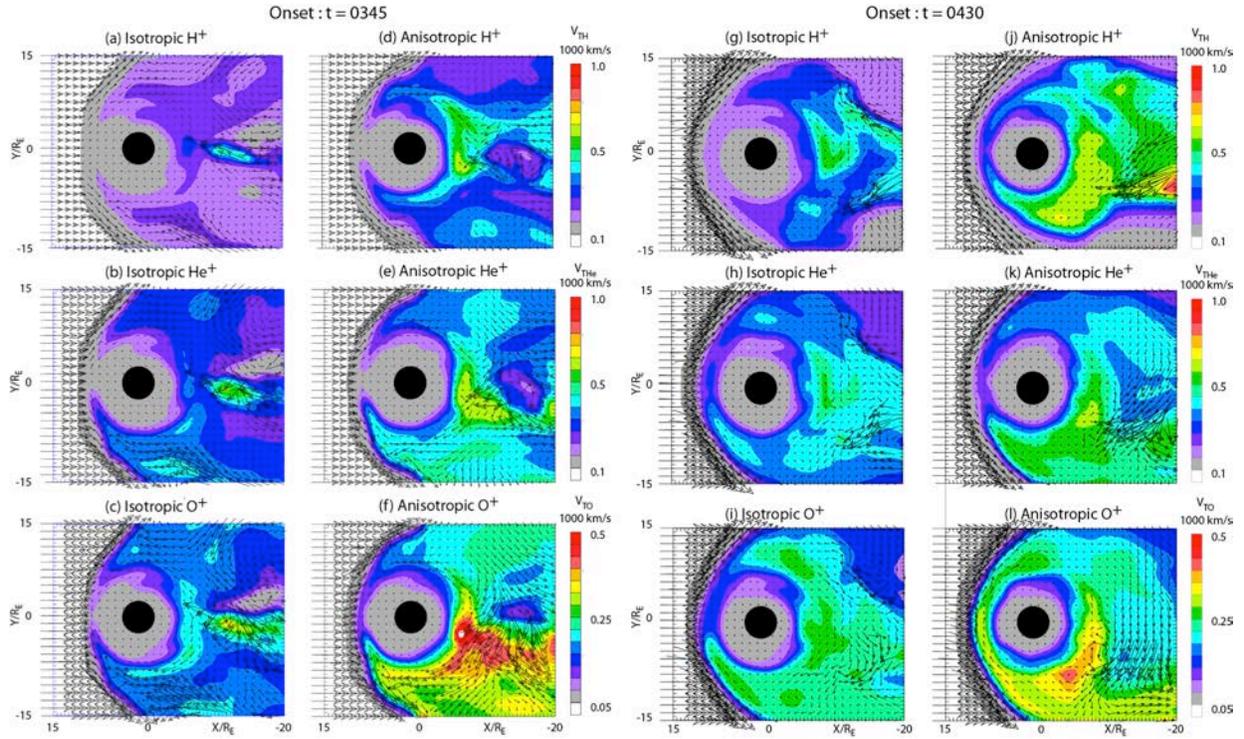

**Figure 11.** The profiles for the equatorial thermal velocity at onset for the $H^+$ (top), $He^+$ (middle) and $O^+$ (bottoms) ions near onset (left) and during the expansion phase (right). In both cases the thermal velocities for the $H^+$ and $He^+$ ions are comparable (i.e. $He^+$ has 4 times the thermal energy of the $H^+$ ions), while the $O^+$ ions attain a thermal speed of about 80% of this speed. The peak value of thermal velocities near onset irrespective of ion species is higher by about 50% for the anisotropic simulations than the isotropic simulations, which leads to a great injection of energetic ions into the inner magnetosphere.

At the actual time of onset, the $O^+$ ions that originated from the cusp/cleft and polar cap are able to reach the plasma sheet for the anisotropic simulations (e.g. Figure 10j). This leads to a stronger concentration of $O^+$ in the lobes and in the current sheet at onset. Note that for both simulations, the density of $O^+$ in the plasma sheet in general remains low throughout most of the expansion phase of the substorm. The variations in the $He^+$ relative density are similar and are not shown.

The overall $O^+$ density in the plasma sheet remains low because the heavy ions are preferentially accelerated relative to the other ions and therefore conservation of mass requires that it have a reduced density in this region [cf. *Winglee et al.* 2008, 2011]. This differential energization of the various ion species is illustrated in Figure 11 which shows contours for the thermal velocity in the equatorial plane at the time of onset with the thermal velocity in the anisotropic simulations being defined as $V_{T_{Av}} = (V_{T_\parallel} + V_{T_{B\times(V\times B)}} + V_{T_{V\times B}})/3$. The left hand side shows the profiles just prior to substorm onset while the right hand side shows the profiles well into the expansion phase.

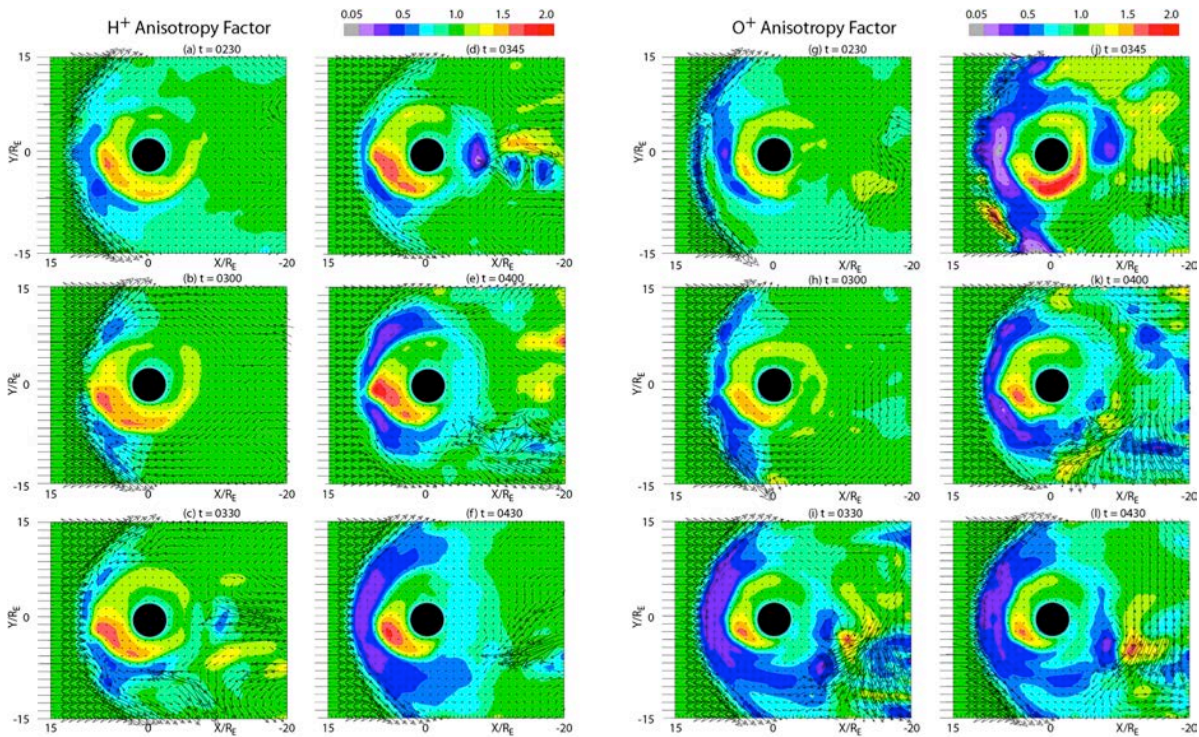

**Figure 12.** Anisotropy factor for the H$^+$ ions (left) and O$^+$ ions (right) over the same region as in Figure 11 and 14 during the development of substorm. The plasmasphere is seen as the population in the inner magnetosphere with $T_\perp < T_\parallel$. During the growth phase, this population is seen to be pulled towards the dayside magnetopause with $T_\parallel / T_\perp$ increasing. This feature is present well after substorm onset. The heavy ions show a higher $T_\perp$ anisotropy on the dayside and a $T_\parallel$ anisotropy tailward of the reconnection region.

The most striking feature is that the anisotropic simulations show that the ions reach much higher overall temperatures than in the isotropic simulations. This effect arises because there has to be a balance between the thermal pressure in the north-south direction and the *J*×*B* force holding the plasma sheet together. In the isotropic simulations any plasma heating automatically increases the thermal pressure which is trying to expand the plasma sheet. In the anisotropic simulations the dawn-dusk electric field is producing acceleration and heating of the heavy ions in the *y*-direction and the light ions experience convective drift in the *x*-direction with the intermediate ions experience a mixture of these two acceleration processes. In all these cases, there is little heating in the *z*-direction so that the ions have a longer residence time in the plasma sheet and experience greater energization in the anisotropic simulations.

For both the isotropic and anisotropic simulations the He$^+$ ions reach comparable thermal velocity's to the H$^+$ ions near local times around midnight (i.e. the He$^+$ ions have about four times the thermal energy). However, there is additional energization of both the ion species on the dusk side as the plasma is convected into stronger magnetic field. This is basically a betatron effect being incorporated in the anisotropic simulations. This preferential heating is even more evident in the anisotropic $O^+$ ions which reach thermal speeds of about 80% of that of the $H^+$ ions near midnight local time (or about 10 times the thermal energy of the light ions). By the time the heavy ions reach the dusk sector they attain thermal speeds comparable with that of the light ions.

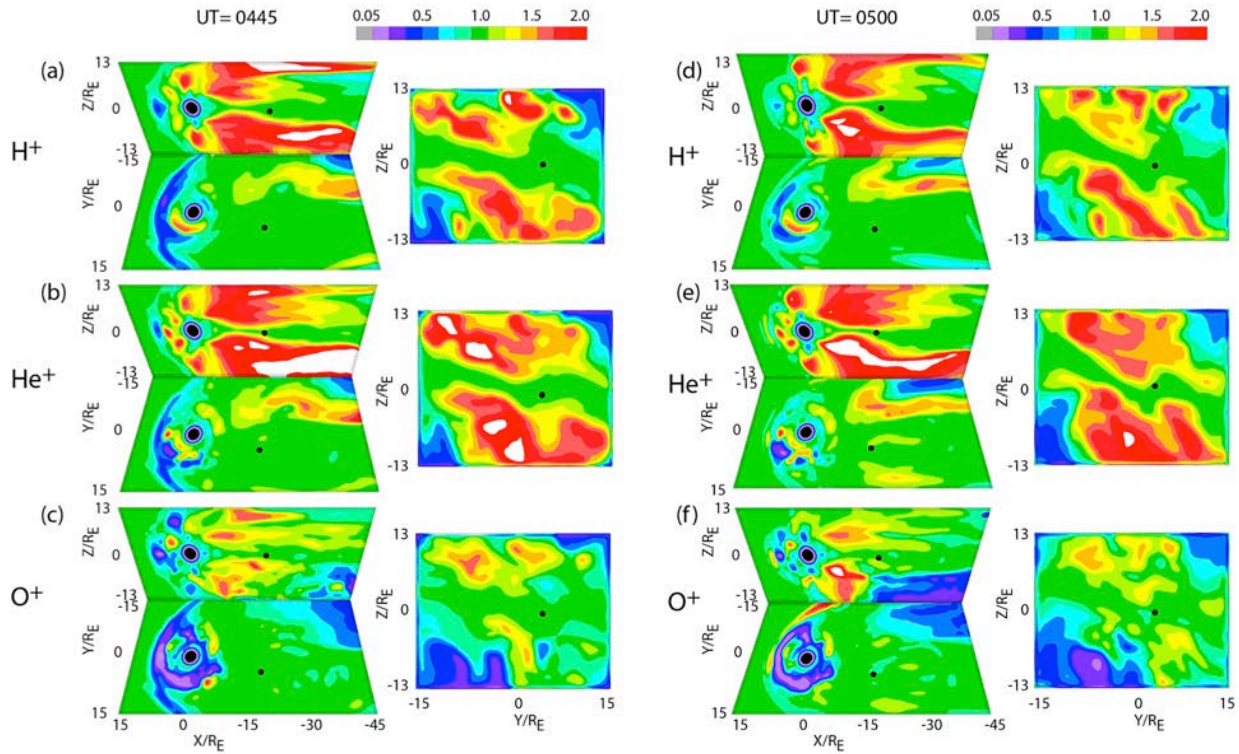

**Figure 13.** The anisotropy factor derived from the model for the 15 September 2001 event. In each panel the anisotropy factor is shown in cuts along the noon midnight meridian and in the equatorial plane plus a cross-section through the average position of cluster which is shown as a black dot. Left-hand side shows the tail configuration just prior to Cluster encountering anisotropic ions in the right hand side shows the configuration as these ions are encountered.

Near the end of the expansion phase (right hand side of Figure 11), the anisotropic simulations indicated continued heating of the ion species as they move into the dusk sector. A crucial difference though between the different ion species is that it is the $O^+$ ions that are primarily able to propagate past noon and contribute to a symmetric ring-current plasma configuration. The drift speed of the other two species is insufficient to overcome the convective electric field near noon which drives these components back down the flanks. This effect has been seen in single particle trajectory analysis of mass and energy transport during substorms and storms [*Cash et al.,* 2010a,b].

The plasma circulating around the inner magnetosphere has distinctive anisotropy factors which can be used to augment the temperature diagnostics and determine how particular regions in the magnetosphere are evolving. As an example Figure 12 shows the anisotropy factor for the $H^+$ ions (left hand side) and $O^+$ ions (right hand side) during the development of a substorm. The cold regions in the inner magnetosphere correspond to the plasmasphere and are seen in Figure 12 to have A > 1 . As the growth phase develops, this plasma component is seen to be pulled towards the dayside magnetopause and $T_\parallel$ increases relative to $T_\perp$. It is only near the end of the expansion phase when the magnetopause has moved outwards under sustained northward IMF does a region of high $T_\perp$ develop between the magnetopause and plasmasphere. Note that the region of high $T_\parallel /$ then convects antisunward. The motion of this feature has similar properties to the plasmaspheric tail identified in the IMAGE observations [*Burch et al.,* 2001].

The corresponding evolution of the anisotropy factor for the $He^+$ ions is similar to that of the $H^+$ ions and is not shown. However, the $O^+$ ions, have some significant differences. The plasmaspheric component is again seen in inner magnetosphere but the anisotropy factor is smaller than for the $H^+$ ions. In addition, a region of high $T_\perp$ develops between the plasmasphere and the dayside magnetopause during the growth phase. The anisotropy factor continues to decrease in this region all the way up through to substorm onset.

In the tail, the $O^+$ ions in entering the nightside plasma sheet are seen to have initially a high $T_\parallel$. However, as they are accelerated duskward and move into the inner magnetosphere, they experience strong perpendicular heating. As a result, the plasma sheet has distinct regions that alternate between high $T_\perp$ and high $T_\parallel$, consistent with the Cluster observations cited in the introduction. As these heated ions move into the dusk sector, they experience additional $T_\perp$ heating so that from pre-midnight all the way through to noon and into the dawn sector A < 1. .

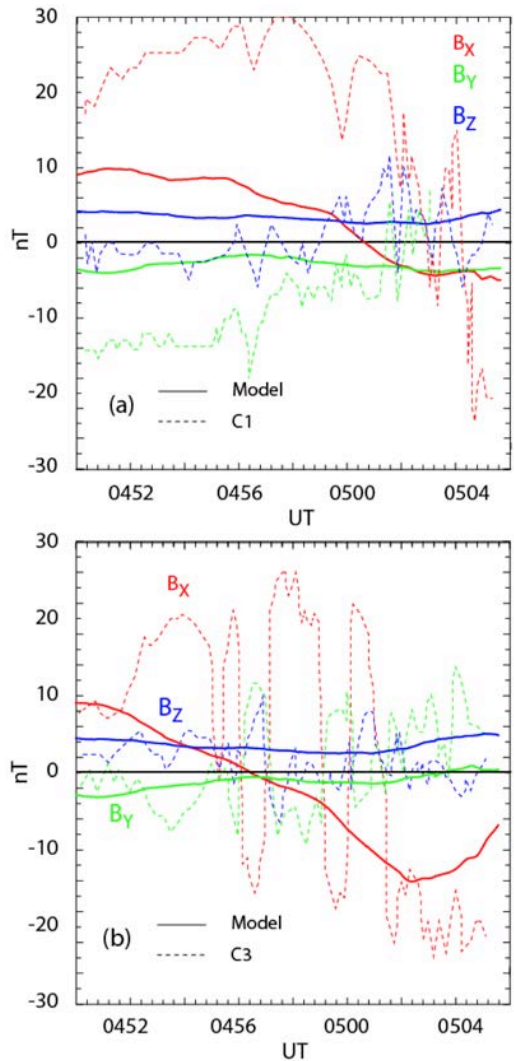

**Figure 14.** Comparison of the perturbations in the magnetic field as derived from the model and is observed by the Cluster C-1 (top ) C-3 (bottom) spacecraft for the event. The C-3 spacecraft with its lower *z* makes a more complete crossing of the hemispheres. This difference in the magnetic field from these different trajectories is seen in the model, though the model under predicts the variance in $B_x$.

## 5. Spacecraft Signatures.

Recently, *Cai et al.* [2008] have shown, using Cluster data, that for a thin current sheet the protons can develop a pressure anisotropy with $T_\perp < T_\parallel$ while the oxygen can develop an anisotropy primarily with $T_\perp > T_\parallel$ (though at the PSBL a region of $T_\perp < T_\parallel$ can be present). Plate 1 shows how the anisotropy factor for H$^+$ and O$^+$ evolve in time at four different positions down the tail. Similar to *Cai et al* [2008], pressures in all three directions are shown with the black line showing the results for the isotropic simulations. Because of the higher temperatures attained, the anisotropic pressures are higher than the isotropic temperatures by a factor of as much as four, though the time variations have similar relative changes between the two simulations. Similar to the results of *Cai et al.* [2008] the two perpendicular temperatures for the protons are approximately the same. At the time of the ejection of the plasmoid (near t = 0330), regions of high $T_\parallel$ are seen at $X \geq -15$ R$_E$ while in the inner magnetosphere a region of high relative $T_\perp$ occurs associated with the breaking of fast flows near the inner edge of the plasmasheet. After onset, the tail plasma becomes approximately isotropic though there are small bursts where there is a $T_\parallel$ excess.

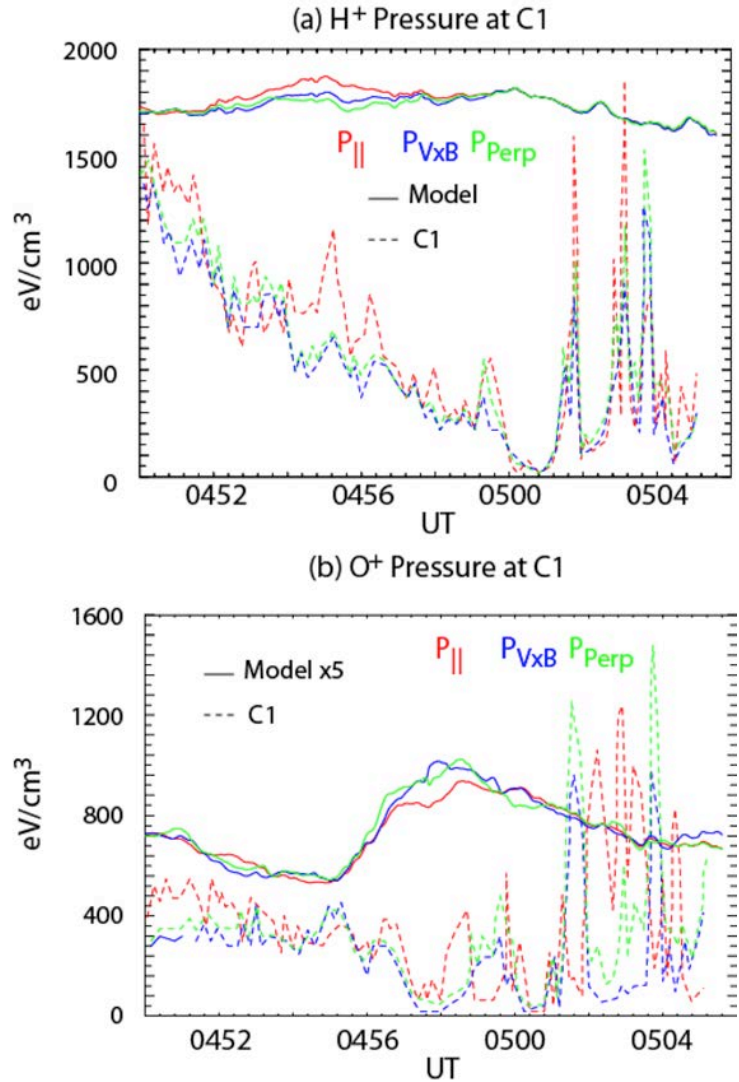

**Figure 15**. Comparison of the temperature anisotropy factors as derived from the model (solid lines) and the C1 spacecraft (dashed lines) for H$^+$ (top) and O$^+$ (bottom). The model has significantly lower O$^+$ concentrations and to put it on the same scale as the observations the model O$^+$ pressure are multiplied by a factor of five. The observations showed significantly more small-scale variations than the model as the center of the current sheet is encountered but overall both the observations and model indicate a slight $T_\parallel$ excess for H$^+$ and a slight $T_\perp$ excess for O$^+$.

For the O$^+$ ions there can be significant differences between the two perpendicular temperatures particularly inside of $X = -15$ R$_E$ where $T_{V \times B} > T_{B \times (V \times B)}$. This difference is strongest during the growth phase and through onset. Such differences in the two perpendicular temperatures are noted in the results of *Cai et al* [2008]. Regions of high relative $T_\perp$ are seen 20-30 minutes prior to onset, though the exact timing is dependent on position down the tail. The perpendicular anisotropy is seen for another 15- 20 min after onset inside of $X = -15$ $R_E$ and out to t ~ 0415 at larger distances down the tail. Regions of high $T_\parallel$ in the tail then appear close to onset and through the expansion phase, which is also consistent with of *Cai et al* [2008].

The magnetic field, which is shown in Plate 2, has quantifiably different behavior between the isotropic and anisotropic simulations. In the inner magnetosphere at $X = -10$ $R_E$ there is substantially more stretching of the magnetic field as seen by the larger $B_X$ prior to substorm onset for the anisotropy simulations. The dipolarization is also very much more abrupt for the anisotropic simulations as seen by the large jumps in both $B_X$ and $B_Z$. The latter magnetic field has a two-step jump during the recovery phase associated with the differences in the development of the Region II currents.

At $X = -15$ $R_E$ and -20 $R_E$, reconnection is seen by the development of negative $B_Z$, which occurs 5 to 10 min earlier than isotropic simulations. In addition, the anisotropic simulations show oscillations with a period of a few minutes in the reconnection region. These oscillations are not present in the isotropic simulations and are associated with ion gyro-motion in the magnetotail. Some of these oscillations have flux ropes embedded in them as seen by the spikes in $B_Y$. The late development of reconnection in the isotropic simulations is also evident at the largest down tail position shown in the figure at $X = -25$ $R_E$. In addition to this timing difference, the flux ropes in the anisotropic simulations are more intense as evidenced by the larger $B_Y$ and $B_Z$ perturbations at this position. Thus, the results presented here indicate that trying to obtain a detailed understanding of substorm development by relative timing of various events needs to be carefully considered because there are additional effects stemming from temperature anisotropies that had yet to be incorporated in most global models.

The difference in the derived timing is highlighted in Plate 3, which shows the $H^+$ ion velocity profiles for the same positions as in Plate 2. At $X = -10$ $R_E$ fast earthward flows are seen a few minutes prior to onset for the anisotropic simulations while the isotropic simulations have these flows delayed by about 10 min, though the peak magnitude of the flow speed is comparable between the two simulations. Both simulations indicated that the flows have a strong dawnward component but within minutes of onset these flows become predominately duskward in the anisotropic simulations. This motion coincides with the westward movement of the nightside auroral currents. In the isotropic simulations the dawnward flow dominates until well into the expansion phase. At $X = -15$ $R_E$ fast tailward flows start to develop at 0315 well before onset for the anisotropic simulations and in front of the isotropic simulations by about 5 mins. About 7 min after onset, bursty bulk flows towards the earth are seen about every 10 mins. The anisotropic simulations show an overall duskward flow while the isotropic simulations indicate a predominantly dawnward component. At larger distances down the tail the isotropic simulations show a reversal in the tail flows about 30-40 min after onset while the anisotropic simulations indicate a more sustained tailward flow at larger distances down the tail.

For the $O^+$ ions (Plate 4), one sees the same types of timing differences between the isotropic and anisotropic simulations in the development of the fast flows. More significantly it is seen that the presence of strong earthward flows is reduced and replaced by strong duskward flows. Thus, the entry of the energetic $O^+$ into the ring currents occurs duskward in the anisotropic simulations relative to that derived in the isotropic simulations. This difference in position is seen in the temperature profiles shown in Figure 12.

In order to provide a closer comparison with in situ observations, the model was rerun with the same inner boundary conditions as the above idealize case but with the solar wind conditions for the period for the magnetotail anisotropy event of 15 September 2001 reported by *Cai et al.* [2008]. The main tail interaction observed by Cluster was around 05 UT. For this event, the IMF $B_z$ was predominantly northward with an average of about +5 nT for about 3 hrs except a small 10 minute southward excursion between 0350 and 0400 UT. IMF $B_y$ was primarily dawnward at about -5 nT. The solar wind density was elevated at 17 $cm^{-3}$ with a speed of 460 km/s between 0230 and 0300 UT. Over next hr, the solar wind density decreased to 6 $cm^{-3}$ and the speed increased to 510 km/s and it remained at these values during the Cluster observations. The mean positions of the Cluster spacecraft was (-18.6, 4.16, -0.9) $R_E$. The C3 spacecraft was about 1500 km down in *z* relative to the other 3 spacecraft. In this section we concentrate on the data from the C1 and C3 spacecraft where the variances in *z* are the strongest.

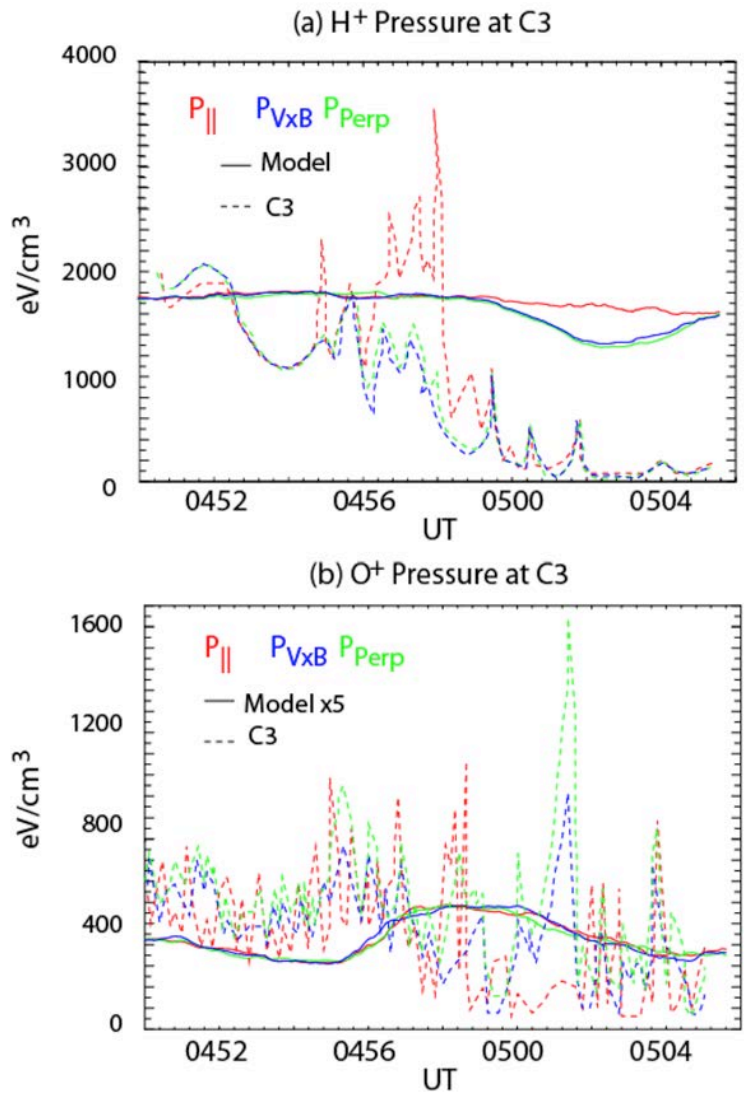

**Figure 16**. As in Figure 15 except for the C3 spacecraft. With the deeper penetration of the C3 spacecraft across the plasma sheet there is stronger $T_\parallel$ excesses in the $H^+$ ions.

Figure 13 shows the anisotropy factor derived from the model just prior to and during Cluster's interaction with the plasma sheet. The average position of Cluster is shown by the black dot in the panels. Because of the relatively strong northward IMF, the anisotropy factors particularly in the lobe are reminiscent of the end of the expansion phase of the idealized case in Figures 6-8. Because of the presence of IMF $B_y$ the plasma sheet is seen to be rotated and kinked in the cross-tail views. The average position of Cluster is predicted to be beyond the thinnest part of the plasma sheet where the strongest anisotropies are present. Nevertheless, a kink in the plasma sheet is seen to move towards Cluster producing an apparent change in the hemisphere in which Cluster resides. As noted in the previous section such as transition is expected to produced variances in the anisotropy factor which should be observable by the spacecraft.

In order to show how the spacecraft crossings look between the in-situ data and the model results, Figure 14 shows the magnetic field profiles as seen by the Cluster spacecraft relative to the derived results from the model. The spacecraft sees significantly more fine scale structure then within the model. This difference arises because in the present work the grid resolution is only 0.15 $R_E$. Another difference is that the lobe fields as seen by the Cluster spacecraft is stronger than the magnetic field seen in the model at the same positions. Nevertheless the model is able to predict the different transitions of the spacecraft between hemisphere with $B_x$ changing sign at about the same time as sign in the Cluster data for both spacecraft.

The corresponding three-component pressures for $H^+$ and $O^+$ at the C1 spacecraft is shown in Figure 15. The model pressures for both ions results show flatter profiles, similar to the magnetic field profiles. Nevertheless the average value of the model results is within a factor of 2 of the observations. The model results also show that on average there should be an excess of in $T_\parallel$ and this excess is see by C1.

The model under predicts the amount of oxygen for quite conditions [cf. Winglee et al., 2008b] so in order to put it on the same scale as observations, the model results are multiplied by a factor of five. The $O^+$ model pressure shows a slight temporary rise in the pressure as the region is encountered. The model results show a slight $T_\perp$ excess associated with this enhanced pressure region. Some $T_\perp$ excess is also seen in the $O^+$ data but at a few minutes later in time. Note that overall both data and model indicate that the behavior of the anisotropy of the $O^+$ ions that is different from the anisotropy of the $H^+$ ions. In comparing the anisotropies for the C3 spacecraft (Figure 16), it is seen that the $T_\parallel$ excess is larger for the $H^+$ ions at the lower $z$ position of the spacecraft. This increase in $T_\parallel$ is associated with light ions that have not fully convected into the plasma sheet at the lower position and experienced additional perpendicular acceleration. This increase in $T_\parallel$ is also seen in the model results. For the $O^+$ ions, there is a lot more noise in the data but overall the $O^+$ ions appear nearly isotropic and this is seen in the model results as well. This result further reinforces the conclusion that ionospheric $H^+$ and $O^+$ ions reach the plasma sheet at different locations and this difference can be directly seen in the local anisotropy factor.

## 6. Summary and Conclusions.

In this paper we have examined the effect of temperature anisotropies on the mass and energy transport within a global magnetospheric model. This model includes for the first time a full treatment of the pressure tensor which overcomes limitations of the double adiabatic treatments where artificial isotropization factors have to be assumed. The present model does not have to make this assumption and it includes for the first time the three major ionospheric species, $H^+$, $He^+$ and $O^+$. The anisotropy factors derived in the present model are shown to comparable to those seen in a quiet time cross of the plasma sheet at 18 $R_E$ by Cluster, which suggests that the present model is not over emphasizing the anisotropy effects. While more effort needs to be made on calibrating the new model, the results are very distinctive in that temperature anisotropy effects can produce:

- A reduction in the saturation value of the cross polar cap potential by between 10 and 20% relative to the value seen in the equivalent isotropic simulation;
- the development of enhanced nightside auroral currents during substorm onset to a value where these currents can exceed the peak value seen on the dayside;
- the ionospheric outflow is seen to develop conic-like features over the polar cap but when these ions enter the lobe the distributions become beam-like with strong parallel temperature anisotropies though the actual value of the temperature anisotropy is dependent on the ion mass and position down the tail;
- an increase in the outflow rate of the light and intermediate ions, particularly around substorm onset while there is little change in the net outflow of the heavy ions;
- the arrival of heavy ions with a large parallel temperature anisotropy leads to faster reconnection rates than in the isotropic simulations so that reconnection occurs earlier;
- the earlier development of reconnection substorm and onset by ~ 10 min earlier (or about 10-20% of the typical growth phase period) than the value obtained from isotropic simulations
- oscillations with a period of a few minutes develop in the reconnection region in the anisotropy simulations (and are not present in the isotropic simulations), and are associated with non-gyrotropic ion distributions (i.e. where all three temperature components are different);
- strong perpendicular heating of the ions on entry into the plasma sheet so that regions of both high $T_\perp$ and high $T_\parallel$ can be present, depending on lifetime of the ions in the plasma sheet;
- the plasmaspheric populations are seen as regions with high $T_\parallel$ and are seen to be pulled towards the dayside magnetosphere during the growth phase; these populations lead to an extended plasmaspheric population in space that eventually convect towards the nightside during the expansion phase;
- for the fully isolated substorm considered here, the $O^+$ ions are primarily the ones that are able to convect across noon LT and these ions can contribute a symmetric ring current.

These results show that anisotropies play a very strong role in determining the dynamics of the global magnetosphere during active periods. The size of the effect increases with mass number of the ion species which is not unexpected as it is these latter ions that can have gyro-radii comparable to scale lengths of boundary layers within the magnetosphere. Future efforts need to quantify the results for storm conditions where the forcing of the magnetosphere is likely to produce even larger anisotropic effects.

**Acknowledgments**

This work was supported by NASA Geospace grant # NNX10AK96G to the University of Washington.

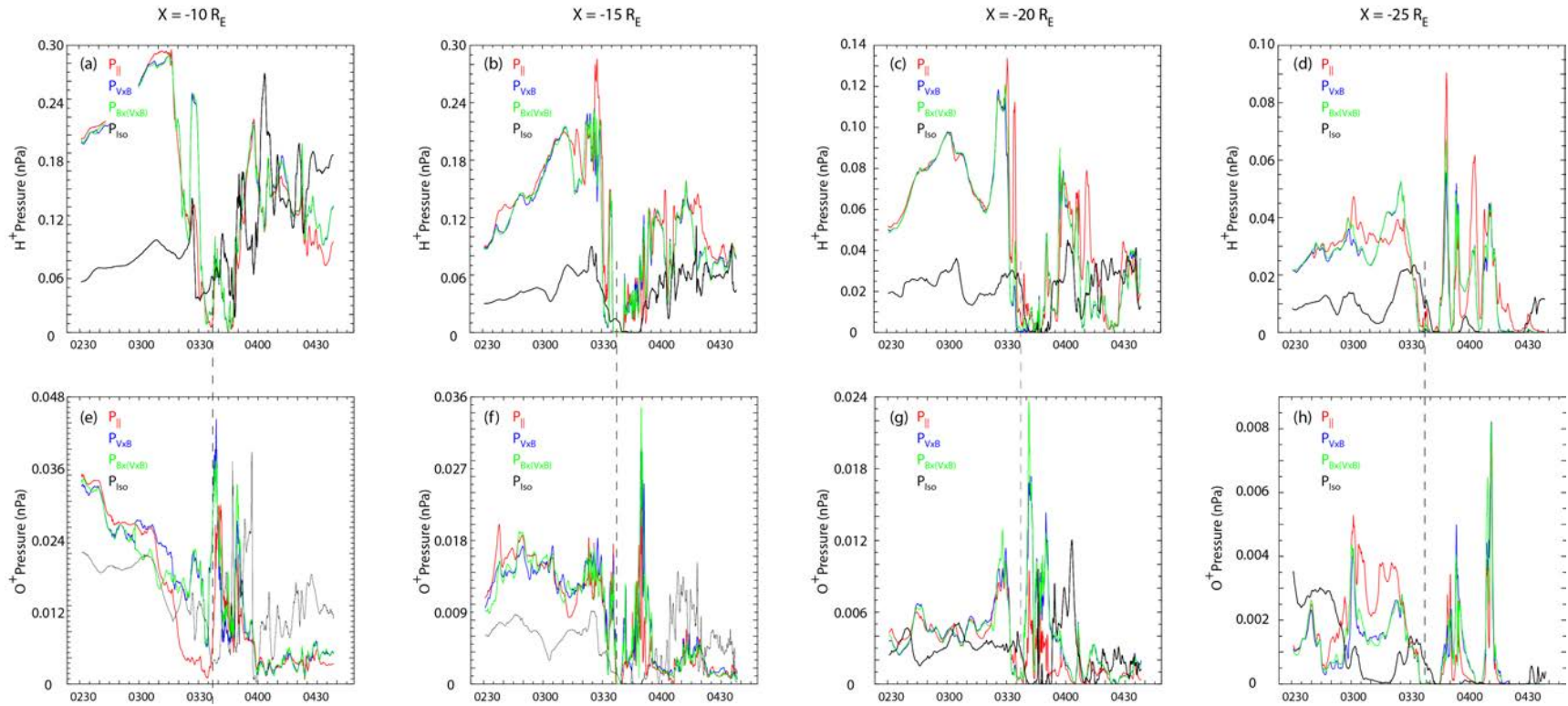

**Plate 1.** Evolution of the H$^+$ (top) and O$^+$ (bottom) pressures at four positions near the center of the current sheet along the noon-midnight meridian. For the H$^+$ ions, the two perpendicular temperatures are almost identical but for the O$^+$ ions there can be substantial differences between the two perpendicular components, particularly near substorm onset and during the expansion phase.

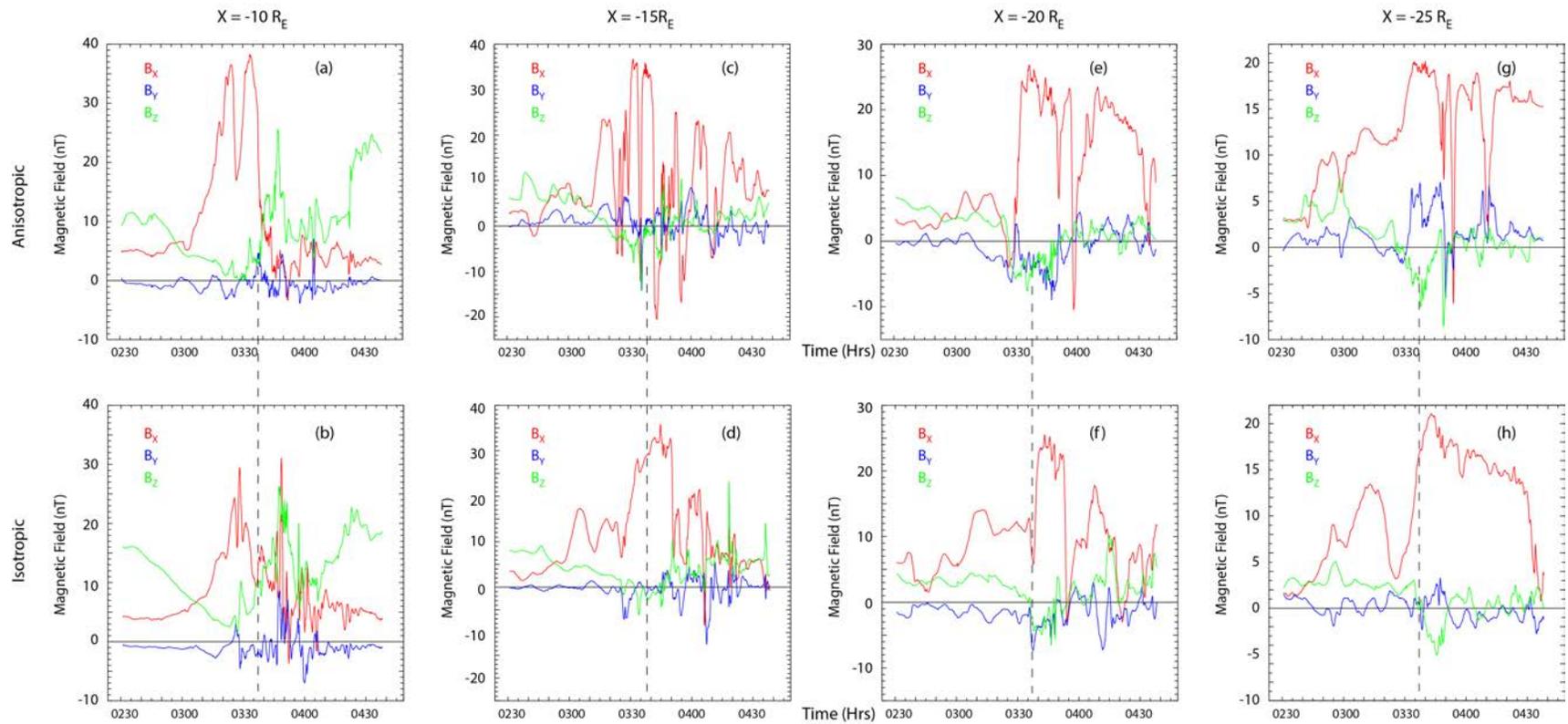

**Plate 2.** Magnetic field perturbations corresponding to Figure 13. Magnetic reconnection in both the isotropic and anisotropic simulations occur prior to onset, with the anisotropic simulations having reconnection occurring first. The anisotropic simulations show more stretching and unloading of the field lines as well as faster oscillations near the reconnection region. at X = -15 $R_E$.

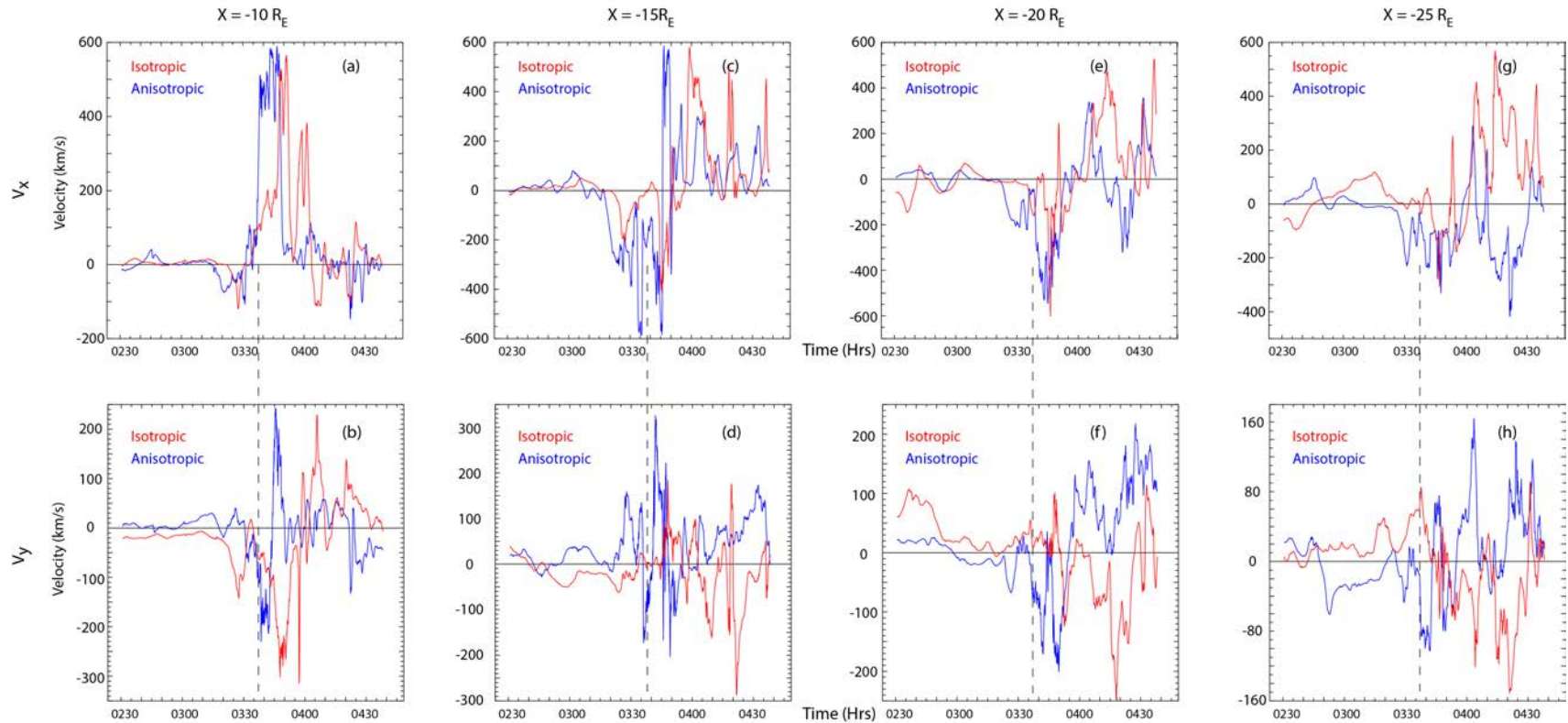

**Plate 3.** Perturbations in $V_x$ and $V_y$ for the $H^+$ ions corresponding to Figure 13. The inclusion of temperature anisotropy effects does not modify the peak speeds attained but they do change the timing and in some places the direction of the flows.

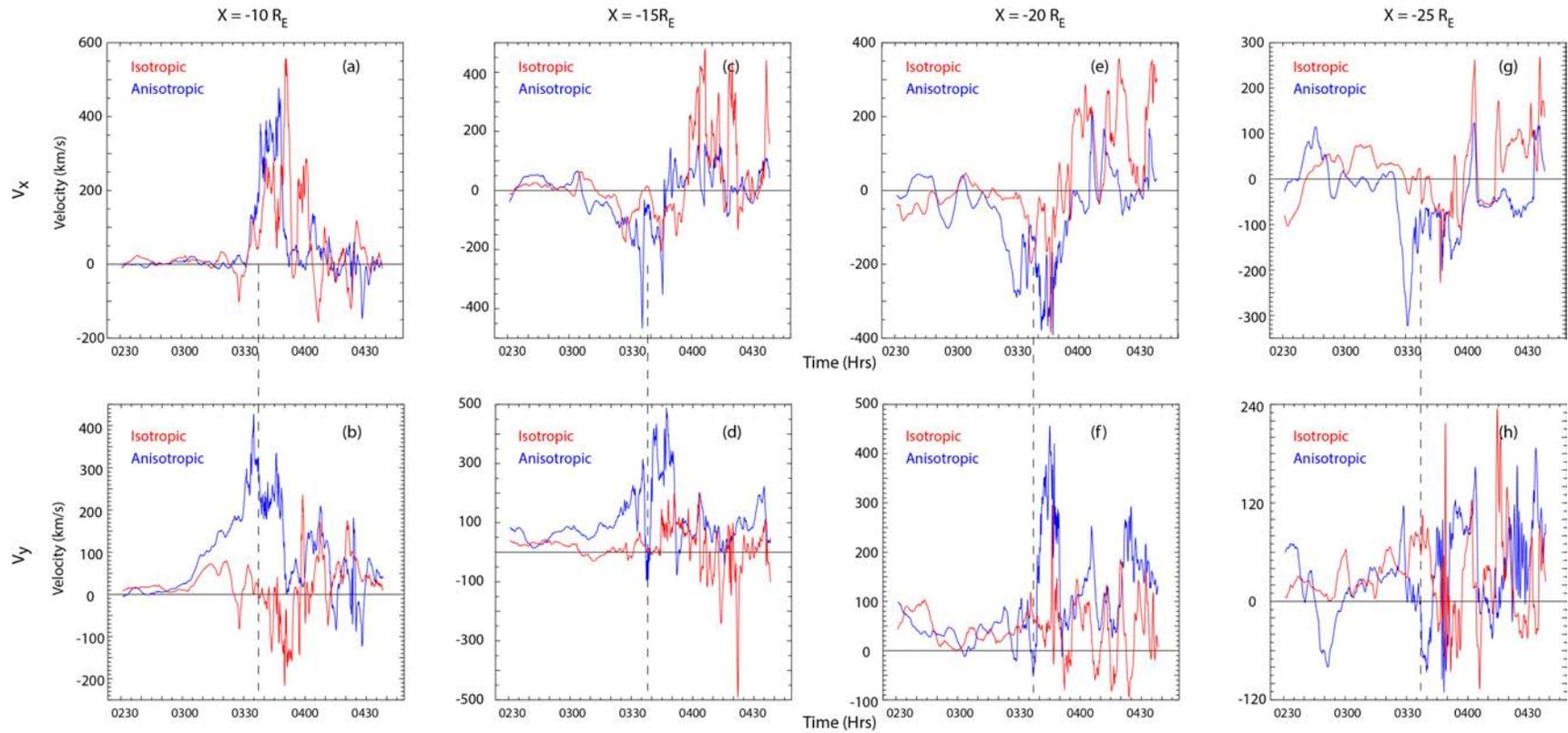

**Plate 4.** As in Figure 15 except for the O⁺ ions. Due to anisotropy effects, the $V_x$ is reduced and $V_y$ is more strongly directed in the duskward direction.